\documentclass[openright, 10pt]{article}
\usepackage{DEF}
\usepackage{DEF-script}
\usepackage{DEF-bold}
\usepackage{dsfont}

\def\z{\mathit{z}}
\def\hz{{\hat{z}}}
\def\hZ{{\hat{Z}}}
\def\Tree{\mathbb{A}}
\def\his{\mathcal{H}}
\def\XX{\mathbf{X}}
\def\WW{\mathbf{W}}
\def\xx{\mathbf{x}}
\def\ww{\mathbf{w}}
\def\TT{\mathbf{T}}
\def\tt{\mathbf{t}}
\def\i{\mathit{i}}
\def\j{\mathit{j}}
\def\l{\mathit{l}}
\def\k{\mathit{k}}
\def\S{\mathit{S}}
\def\s{\mathit{s}}
\def\C{\mathit{C}}
\def\c{\mathit{c}}
\def\pro{{^*}}
\def\root{{0}}

\def\Xpp{{\Xi}}
\def\Ypp{{\Upsilon}}

\def\xpp{{\xi}}
\def\ypp{{y}}

\def\Xp{{X}}
\def\Yp{{Y}}
\def\T{{T}}
\def\xp{{x}}
\def\yp{{y}}
\def\t{{t}}
\def\u{{u}}

\def\ch{{\rm ch}}
\def\pa{{\rm pa}}
\def\an{{\rm an}}
\def\tmin{t_{\rm min}}
\def\tmax{t_{\rm max}}
\def\tobs{t_{\rm obs}}
\def\ts{t_{\rm syn}}

\def\lik{\ell}

\def\ker{{K}}
\def\P{{P}}
\def\Q{{Q}}

\def\p{{p}}

\def\det{{G}}

\def\I{\mathcal{V}}
\def\Iu{{\I_{\rm act}}}
\def\It{{\I_{\rm end}}}
\def\L{\mathbb{L}}

\title{Particle MCMC with Poisson Resampling:\\
       Parallelization and Continuous Time Models}
\author{T. C{a}ka{l}a,  B. Miasojedow, W. Niemiro}

\begin{document}

\maketitle

\begin{abstract}
We introduce a new version of particle filter in which the number of ``children'' of a particle
at a given time has a Poisson distribution. As a result, the number of particles is random and varies with time.
An advantage of this scheme is that descendants of different particles can evolve independently. 
It makes easy to parallelize computations. Moreover, particle filter with Poisson resampling is readily adapted to
the case when a hidden process is a continuous time, piecewise deterministic semi-Markov process.
We  show that the basic techniques of particle MCMC, namely 
particle independent Metropolis-Hastings, particle Gibbs Sampler and its version with ancestor sampling, 
work under our Poisson resampling scheme. Our version of particle Gibbs Sampler 
is  uniformly ergodic under the same assumptions as its standard counterpart. We present simulation results
which indicate that our algorithms can compete with the existing methods.
\end{abstract}


\section{Introduction}\label{sec:Intro}
Particle Filters (PF) and more generally Sequential Monte Carlo methods (SMC) \citep{Gordon1993, Doucet2001, Moral} are general framework 
for statistical inference for state space models. SMC methods have proven effective in various scenarios covering: object tracking, time series analysis in 
non-Gaussian models \citep{Gordon1993, Doucet2001}, graphical models \citep{naesseth2014sequential}, rare events estimation \citep{Cerou2011},
phylogenetic inference \citep{Bouchard-Cote2012} and model selection \citep{Schaefer2011}. 
The seminal paper \citep{Andrieu2010} introduced Particle Markov Chain Monte Carlo methods (PMCMC), which combine strengths of
MCMC and SMC algorithms.  

Most of the research on SMC methods and their extensions is focused on discrete time models. Statistical inference for continuous time models is usually
performed via discretisation of time, see for example \citep{LotkaVolterr} and by introducing rather complex birth-death moves \citep{Finkle}.
The main difficulty in designing SMC methods for continuous time models is the fact that the standard resampling step requires synchronisation of all the paths.
In the current paper we introduce  a unified approach for both discrete and continuous time setting.
We propose a new Poisson resampling scheme. Rather surprisingly, this scheme allows for a straightforward extension of PMCMC to a 
wide class of piecewise deterministic processes (PDP). They are processes which evolve deterministically
in continuous time except for a countable collection of stopping times at which they randomly jump, see \citep{Davis1984}. 
PDPs have recently attracted much attention because they are  most natural models of a lot of phenomena in biology
\citep{Rudnicki2017} and in other branches of science.

In addition, the standard resampling scheme is challenging to implement  in parallel, see for example \citep{Paige2014,Murray2016}.   
Our scheme is much easier to parallelise, 
due to the fact that only partial synchronisation is required in Poisson resampling.
Our algorithm PTPF (Poisson Tree PF), similarly to \citep{Paige2014} generates a branching process. 
The main advantage of our approach, and the difference from \citep{Paige2014}, is that PTPF can be directly used within 
PMCMC methods. Moreover, our framework allows us to perform ancestor sampling in the particle Gibbs algorithm 
in the spirit of \citep{Lindsten2014}. We prove that our version of Particle Gibbs Sampler (in the discrete time setting)  
is  uniformly ergodic, under the same assumptions as its standard counterpart \citep{LindDoucMoul2015}.

Since the Poisson resampling produces a random (and varying) number of particles, it is essential to control their population size.
To make our algorithms  practically applicable, we have to introduce some sort of synchronisation of particles. 
(Even though synchronisation is \textit{not} necessary to ensure convergence to the target.) The inherently parallel structure 
of Poisson resampling has to be reconciled with (partial) synchronisation. 
This is relatively easy for discrete time models and much harder for continuous time models. Nonetheless,
we have designed some rules (recipes for choosing control parameters in PTPF) which keep the population of particles approximately
constant.

Apart from theoretical considerations,
we demonstrate our method on a few challenging examples. Our simulations indicate that in the discrete time setting 
our algorithms yied results as good as standard PMCMC. Parallel implementation of our algorithms leads to significant gains in efficiency.  
For continuous time models our algorithms can compete with the existing methods, and in some examples outperform them.

The paper is organised as follows. In Section~\ref{sec:SSM} we introduce a general class of semi-Markov state space models including discrete time 
state space models and piece-wise deterministic hidden Markov processes. Next in the Section~\ref{sec:PoissTreePF} we present
our basic Particle Filter (PTPF).
In Section~\ref{sec:PMCMC} we show how to construct PMCMC methods based on PTPF. In Section~\ref{sec:Variants} we 
introduce specific rules designed to control the size of the population of particles and we define our version of 
ancestor resampling. Finally, in Section~\ref{sec:Sim} we present numerical simulations. 

\section{Semi-Markov State-Space Models}\label{sec:SSM}

We consider a rather general family of state-space 
semi-Markov models. 
We first consider continuous time models.  
Assume that $\{\Xpp(\t),\t\geq\tmin\}$ is a \textit{piece-wise deterministic process} with values in a Polish space $\X$
and with c\`adl\`ag trajectories. The process jumps randomly at a countable set of random times $\T_1<\cdots<\T_k<\cdots$.  Between jumps it evolves 
according to a deterministic law.  We consider a finite time horizon $\tmax$ and write $M=\min\{k:\T_k>\tmax\}$. 
Assume that process $\Xpp=\{\Xpp(\t),\tmin\leq\t\leq\tmax\}$ is uniquely determined by its space-time \textit{skeleton} $(\Xp_{1:M},\T_{1:M})$,
where
\begin{equation}\label{eq:EndPoint}
 \Xp_{k}=\Xpp(T_{k}-),\qquad k=1,\ldots,M.
\end{equation}
Note that our definition of the skeleton is different from the standard one, e.g.\ in \cite{Davis1984}, and perhaps less intuitive.
We require that the trajectory of $\Xpp$ in the time interval $[\T_{k-1},\T_{k}[$ depends deterministically on
its value at the \textit{end} of the interval. (We define the skeleton via \eqref{eq:EndPoint} 
to facilitate our construction of ancestor sampling in Section \ref{sec:Variants}.) We assume that random
variable $M$ is almost surely finite.
Our basic assumption is that the skeleton  is
Markovian, governed by a space-time stochastic transition kernel $\ker$. The (prior) probability distribution of the skeleton 
can be written in a concise form
\begin{equation}\label{eq:Prior}
\begin{split}
 \pi_{\rm prior}(\d \xp_{1:m},\d \t_{1:m})
 &=\Pr(\Xp_1\in\d\xp_1,\T_1\in\d \t_1,\ldots,\Xp_m\in\d\xp_m,\T_m\in\d\t_m)\\
 &=\prod_{k=1}^{m} \ker(\xp_{k-1},\t_{k-1};\d\xp_{k},\d \t_{k}),
\end{split}
\end{equation}
if we adopt a convention explained below. The initial  distribution is written as 
$\Pr(\Xp_1\in\d\xp_1,\T_1\in\d \t_1)=\ker(\xp_0,\t_0;\d\xp_1,\d\t_1)$, where $(\xp_0,\t_0)$ is a ficticious state with $\t_0=\tmin$. 
Note also that the last point of the skeleton $(\xp_m,\t_m)$ falls beyond the time horizon ($\t_m>\tmax$).
In the sequel, $\xpp$ denotes a sample path of $\Xpp$ and we write  $\xpp_{[\t',\t''[}=\{\xpp(\t), \t\leq\t< \t'\}$. 
{Some more details and an explicit construction of $\Xpp$ are in the Supplementary Material.}   

The setup described above encompasses continuous time \textit{piece-wise deterministic Markov processes} \citep{Davis1984},
in particular pure jump Markov processes, and a wide class of piece-wise deterministic non-Markovian processes
\citep{Whiteley, Finkle}. 

The process $\Xpp$ is hidden and thus \eqref{eq:Prior} plays the role of the prior. 
Let $\Ypp$ be an observed random element which depends on $\Xpp$.
The \textit{target probability distribution} is the posterior of $\Xpp$ given 
$\Ypp=\yp$. Since $\ypp$ is fixed, it will not be explicitly indicated. We only need to 
assume that we have a family of likelihood functions $\lik(\xpp_{[\t,\t'[})$ which satisfy the condition
\begin{equation}\label{eq:LikBasic}
 \lik(\xpp_{[\t,\t''[})=\lik(\xpp_{[\t,\t'[})\lik(\xpp_{[\t',\t''[}),
\end{equation}
for $t<t'<t''$.  (By convention, $\lik(\xpp_{[\t,\t'[})$ is understood as $\lik(\xpp_{[\t,\tmax]})$ whenever $t'>\tmax$.)
In most applications the likelihoods satisfy \eqref{eq:LikBasic}. First typical example is when
$\Ypp=(\Yp_1,\ldots,\Yp_p)$ is just a sequence of ``noisy measurements'' on $\xpp$ at
discrete ``observation times'', say $\tmin\leq\tobs^1<\cdots<\tobs^p\leq\tmax$. We assume that $\Yp_r$ depends only on $\xpp(\tobs^r)$
and $\lik(\xpp_{[\t,\t'[})$ corresponds to $\{\yp_r: \tobs^r\in[\t,\t'[\}$. The second example is when 
$\Ypp$ a fully observed contionuous time random process $\{\Ypp(t):\tmin\leq\t\leq\tmax\}$ and $\lik(\xpp_{[\t,\t'[})$ corresponds to $\yp_{[\t,\t'[}$.

Recall that $\xpp_{[\tmin,\tmax]}$ is represented by its skeleton $(\xp_{1:m},\t_{1:m})$. 
Since the trajectory $\xpp_{[\t_{k-1},\t_{k}[}$   depends  deterministically on $\xpp(\t_{k}-)=\xp_k$, we can write 
$\lik(\xpp_{[\t_{k-1},\t_{k}[})=\lik(\xp_{k};\t_{k-1},\t_{k})$. The posterior distribution of  $(X_{1:M},T_{1:M})$ is given by
\begin{equation}\label{eq:Posterior}
\begin{split}
 \pi_{\rm post}(\d\xp_{1:m},\d \t_{1:m})&=\frac{1}{\z}\cdot
                         \prod_{k=1}^{m}\ker(x_{k-1},\t_{k-1};\d x_{k},\d\t_{k})\lik(\xp_{k};\t_{k-1},\t_{k}),
\end{split}
\end{equation}
where $\z$ is a norming constant (the integral of the likelihood with respect to the prior). 
Our main objects of interest are $\z$ and the posterior $\pi=\pi_{\rm post}$.   
From now on, we most often drop the subscript `post'.   

Discrete time hidden Markov models fit in our setup as a special case (identified with piece-wise constant processes).
Let $\Xpp=(\Xp_1,\ldots,\Xp_m)$ be a discrete time Markov chain (in general, inhomogeneous in time) with one-step
transition kernels $\P_1,\ldots,P_{m-1}$. Using a convention explained earlier, the joint (prior) probability distribution is 
\begin{equation}\nonumber
\begin{split}
\pi_{\rm prior}(\d\xp_{1:m})&=\Pr(\Xp_1\in \d\xp_1,\ldots,\Xp_m\in\d\xp_m)=\prod_{\t=1}^m\P_{\t-1}(\xp_{\t-1},\d\xp_\t).
\end{split}
\end{equation}
The natural assumption about the process of observations in the discrete time  setting is that $\Ypp=(\Yp_1,\ldots,\Yp_m)$, where $\Yp_t$ 
depends only on \textit{one} state $\Xp_{t}$ of the Markov chain.
The likelihood is of the form $\lik_\t(\xp_{\t})$ and consequently, 
\begin{equation}\label{eq:PosteriorDiscr}
\begin{split}
 \pi_{\rm post}(\d\xp_{1:m})&=\frac{1}{\z}\cdot
                         \prod_{t=1}^{m}\P_{t-1}(x_{t-1},\d x_{t})\lik_t(\xp_{t}).
\end{split}
\end{equation}

\section{Poisson Tree Particle Filter}\label{sec:PoissTreePF}

To define \textit{Poisson Tree Particle Filter} (PTPF) and particle MCMC algorithms based on PTPF 
we introduce suitable notations. PTPF produces a random structure $\Tree=(\I,\E,\XX,\TT,\S)$.
\begin{itemize}
 \item $(\I,\E)$ is a directed graph with the set $\I$ of nodes and set $\E$ of edges (arrows). 
  
 \item $\XX=\{X_{\i}: \i\in\I\}$ is a collection of random variables  with values in 
  $\X$. 

 \item $\TT=\{T_{\i}: \i\in\I\}$ is a collection of random variables  with values in 
  $[\tmin,\infty[$.     
 
 \item $\S\in\I$ is a (random) node identifying a selected path in the graph.
\end{itemize}
We will also consider two collections of random variables $\LLambda=\{\Lambda_{\i}: \i\in\I\}$ and 
$\WW=\{W_{i}: \i \in\I\}$, which are functions of $\Tree$ (and of the fixed observation $\Ypp=\ypp$).  

Graph $(\I,\E)$ is a directed forest. Every node has at most one incoming edge. 
A generic element of $\I$ is denoted by $\i$ and a generic element of $\E$ by $\i\to\j$. 
If $i\to\j$ then we write $\i=\pa(\j)$ and $\j\in\ch(\i)$. 
It is convenient to add a fictitious node $\root$ to $\I$ and treat the graph as a tree with root $\root$, 
adding arrows $\root\to \i$ for all nodes $\i$ with $\pa(\i)=\emptyset$.  
For any $\i\in\I$ there is a unique ancestry line denoted by $\an(\i)$. It is a sequence of nodes 
$(a_1(\i),\ldots,a_{k}(\i))$ such that $a_k(i)=i$, $a_{r}(\i)=\pa(a_{r+1}(\i))$ for $r=1,\ldots,k-1$ and 
$\pa(a_1(\i))=\root$ (note that $\an(\i)$ includes $\i$ and does not include the artificial root $\root$).
We also write $X_{\an(\i)}=(X_{a_1(\i)},\ldots,X_{a_k(\i)})$ and  $T_{\an(\i)}=(T_{a_1(\i)},\ldots,T_{a_k(\i)})$.     
To every $\i\in\I$ there corresponds a sample path of continuous time process $\Xpp_\i=\{\Xpp_\i(t): \tmin \leq t<T_\i\}$ 
determined by the space-time skeleton $(X_{\an(\i)},T_{\an(\i)})$ (note that   $\Xpp_\i$ is defined on the right open interval
$[\tmin,T_\i[$ and $X_{\i}=\Xpp(T_{\i}-)$, in accordance with the conventions introduced in the previous section). 
 
We first  describe PTPF informally and explain the role played by all the 
involved variables. Let us think that node $\i$ (or equivalently edge $\pa(\i)\to\i$) is an identifier of a ``particle'' 
which is born at time $T_{\pa(\i)}$. Particle $\i$  evolves deterministically from its initial location
till time $T_{\i}$ and $X_{\i}$ denotes its location immediately prior to $T_{\i}$.
If $T_{\i}> \tmax$ then we say $\i$ is a terminal node, $\i\in\It$. Otherwise, $\i$ gives birth to a set $\ch(\i)$ of children. 
This is done as follows. First we choose an ``intensity parameter'' $\Lambda_{\i}$ (see the paragraph below). 
We compute the weight $W_{\i}$  equal to $\lik(X_{\i};T_{\pa(\i)},T_{\i})$, i.e.\ the likelihood corresponding to the deterministic
part of trajectory in the interval
$[T_{\pa(\i)},T_{\i}[$. Then we sample $N_{\i}\sim \poi(\Lambda_{\i} W_{\i})$ and create a set $\ch(\i)$ of cardinality
$N_{\i}$ (possibly empty) with arrows from $\i$ to all $\j\in\ch(\i)$. For every child $\j\in\ch(\i)$ we independently
sample random pair $(X_{\j},T_{\j})$ from the probability distribution $\ker(X_{\i},T_{\i};\cdot,\cdot)$.
Every child $\j\in\ch(\i)$ immediately jumps to its initial location and evolves deterministically till time $T_{\j}$. 
This procedure is repeated until no ``active'' nodes are left. A node $\i$ is said to be active, $\i\in\Iu$, if $T_\i\leq \tmax$
and it has not yet undergone the ``propagation procedure'' described above. 
The last stage of PTPF is selecting one node $\S$ among nodes $\i$ which satisfy $T_\i>\tmax$. The ancestry line $\an(S)$ identifies
a sample path of the hidden process $\{\Xpp(t), t\in[\tmin,\tmax]\}$ which is used as an update in pMCMC algorithms.
We also compute an estimate $\hZ$ of the norming constant $\z$.       

A few more notations are needed to define PTPF more precisely. 
Assume that for any active node $\i$, the corresponding intensity parameter $\Lambda_{\i}$ can depend on the history
of the whole process before the current time $T_{\i}$.  To avoid vicious circle, 
at every stage we can pick up (for ``propagation'') an active node $\i$ with the least $T_{\i}$.
(This last rule is introduced to simplify presentation. Later, in Section \ref{sec:Variants}, it will be relaxed.) 
History up to time $T_\i$, denoted $\his(T_\i)$, is defined as 
a subtree which includes nodes $\l$, $\j$ and  arrows $l\to\j$ such that $T_\l<T_\i$, 
together with the corresponding variables $X_\l,X_\j$, $T_\l,T_\j$ (let us remember that $X_\j$ 
determines the location of a particle born at moment $T_\l$). 
In other words, $\his(T_\i)$ contains information about all the particles $\j$
born before $T_\i$ and allows us to compute the likelihoods $\lik(\Xpp_{\j [t',t''[})$ for $\tmin\leq t'<t''\leq T_\j$.
Every parameter $\Lambda_\i$ is a function of the history, say $\Lambda_\i=\L(\his(T_\i))$.
Some concrete forms of function $\L$ will be discussed in Section \ref{sec:Variants}.
The initial $\Lambda_\root$ is equal to a constant $\lambda_\root$ chosen \textit{a priori}. 
A pseudo-code defining PTPF is the following.

\begin{center}
 Algorithm PTPF (Poisson Tree Particle Filter)
\end{center}\smallskip\nobreak
\hrule
\smallskip\nobreak
\algsetup{indent=3em}
\begin{algorithmic}
    \STATE \COMMENT{ \blu{Initialize:} } 
    \STATE $\I:=\Iu:=\{\root\}$;  $\E:=\emptyset$; $\It:=\emptyset$; $T_\root:=t_0$; $C_\root:=\Lambda_\root:=\lambda_0$; $W_\root:=1$                   
    \STATE \COMMENT{ \blu{Main loop:} }
    \WHILE {$\Iu\not=\emptyset$}
    \STATE Choose $\i\in\Iu$ with minimum $T_\i$ \; \COMMENT{\red{This requirement will be relaxed}}
    \IF {$\i\not=\root$} 
    \STATE Compute $\Lambda_\i:=\L(\his(T_\i))$ \: \COMMENT{\red{This step will be precised later}}
    \STATE $\C_i:=\C_{\pa(\i)}\Lambda_i$  
    \ENDIF 
    \STATE Sample $N_{\i}\sim \poi(\Lambda_{i}W_\i)$
    \IF {$N_{\i}>0$}
    \STATE Create set $\ch(\i)$ of cardinality $N_\i$
    \STATE $\I:=\I\cup\ch(\i)$, $\E:=\E\cup\{\i\to\j: \j\in\ch(\i)\}$
    \FORALL  {$\j\in\ch(\i)$}
    \STATE Sample $(X_\j,T_\j)\sim \ker(X_\i,T_\i;\cdot,\cdot)$  \: \COMMENT{\blu{ Propagate}}
    \STATE Compute $W_\j:=\lik(X_{\j};T_{\i},T_\j)$  \: \COMMENT{\blu{ Weigh }}
    \IF {$T_j>\tmax$} 
    \STATE $\It:=\It\cup\{\j\}$
    \ELSE 
    \STATE $\Iu:=\Iu\cup\{\j\}$
    \ENDIF
    \ENDFOR 
    \ENDIF
    \STATE $\Iu:=\Iu\setminus\{\i\}$ 
    \ENDWHILE 
    \STATE  \COMMENT{ \blu{Select $S$:} } 
    \IF {$\It\not=\emptyset$} 
    \STATE $\hZ:=\displaystyle\sum\limits_{\i\in\It}{W_{\i}}/{\C_{\pa(i)}}$
    \STATE Select $\S\in\It$ from the probability distribution  $\Pr(\S=\s)\propto {W_{\s}}/{\C_{\pa(s)}}$ 
    \ELSE
    \STATE $\hZ:=0$
    \ENDIF
    \STATE Output $\hZ$, $(X_{\an(\S)},T_{\an(\S)})$ \quad \COMMENT{ \blu{Optionally $\Tree=(\I,\E,\XX,\TT,\S)$ } }
    \end{algorithmic}
\nobreak\hrule

For discrete time models, with $\Xpp=(\Xp_1,\ldots,\Xp_m)$ a few details in PTPF become simpler. 
We can omit $\T_{1:m}$ in the input/output. The tree produced by the algorithm
is uniquely represented by $(\I,\E,\XX,\S)$. Kernel $\ker(x_\i,t_\i,\d x_\j,\d t_\j)$ is reduced to $\P_{\t-1}(x_\i,\d x_\j)$, where
$t_\i=\t-1$ and $t_\j=\t$. The set of nodes is partitioned into ``generations'' $\I_t=\{\i\in\I: T_\i=t\}$, $t=1,\ldots,m$.
Nodes belonging to $\I_t$ propagate simultaneously and independently. The set of terminal nodes is $\It=\I_{m}$.

\subsection{Extended probability distributions} 

The joint probability distribution of all the random variables in \[\Tree=(\I,\E,\XX,\TT,\S)\] is called the 
\textit{extended proposal}, following the terminology established in the SMC literature. 
The extended proposal is denoted by $\psi(\I,\E,\d\xx,\d\tt,\s)$.
Values of random variables $X_\i$, $T_\i$ and $\S$  are denoted by the corresponding small 
case letters $x_\i$, $t_\i$ and $\s$. Analogously, notations   $\lambda_\i$, $w_\i$ and $\hz$ will be used
for values of random variables  $\Lambda_\i$, $W_\i$ and $\hZ$, which are functions 
of $(\I,\E,\XX,\TT)$. Consequently, in the formulae below we use the following notations.
\begin{equation}\nonumber
\begin{split}
 & w_\root=1,\qquad   w_\i=\lik(x_{\i};t_{\pa(\i)},t_\i),\qquad \lambda_\i=\L(\his(t_\i)),\\
 & \It=\{\j\in\I: t_\j>\tmax\},\qquad  c_\j   =\lambda_0\prod_{\i\in\an(\j)} \lambda_i.
  \end{split}
\end{equation} 

\begin{rem}[Equivalence classes]\label{rem:labels}
The labels given to nodes of the graph $(\I,\E)$ are irrelevant to the behaviour of the algorithm.  Strictly speaking, we are interested in 
the \textit{equivalence classes}  $[\Tree]=[(\I,\E,\XX,\TT,\S)]$, where two structures are equivalent if they differ from each other
only by labelling of the nodes.  (That is, if there is a one-to-one correspondence
between the sets of nodes which preserves the set of arrows, the variables $X_\i$ $T_i$ and $\S$.) 
In a single ``propagation'' step of PTPF, node $\i$ ``produces'' $n_\i$ children with probability
\begin{equation}\nonumber
 \exp[-\lambda_\i w_\i]\dfrac{(\lambda_\i w_\i)^{n_\i}}{n_\i!}.
\end{equation}
A child with label $\j$ is then assigned a pair $(\xp_\j,\t_\j)$ drawn from $\ker$.
There are $n_\i!$ equivalent configurations of children. Therefore, if we consider the distribution of the {equivalence class},
then the factorial in the Poisson probability cancels out. Let us introduce the following convention.   
From now on, \textit{we work with the equivalence classes} without making explicit the  distinction between a class 
$[\Tree]$ and its representative $\Tree$.
\end{rem}

Now we are in a position to write a formula for the extended proposal.
It is convenient to discern two stages: first the marginal distribution of all the variables except $\S$,
and then the conditional distribution of $\S$ given the rest. This exactly corresponds to the two stages of PTPF: in the ``Main loop''
we sample  $(\I,\E,\XX,\TT)$ and the last part of the algorithm is ``Selecting $S$''. 

The \textit{extended proposal} is given by
\begin{equation}\label{eq:GenExProp}
\begin{split}
  &\psi(\I,\E,\d\xx,\d\tt)= 
  \prod_{\i\in\I\setminus\It} \exp\left[-\lambda_{\i}w_{\i}\right] \left(\lambda_{\i}w_{\i}\right)^{|\ch(\i)|}
  \prod_{\j\in\ch(\i)}\ker\left(x_{\i},t_{\i};\d x_{\j},\d t_{\j}\right);\\
  &\psi(\I,\E,\d\xx,\d\tt,\s)= \psi(\I,\E,\d\xx,\d\tt)\frac{w_{\s}/\c_{\pa(\s)}}{\hat{z}},
\end{split}
\end{equation}
where
\begin{equation}\nonumber
 \hat{z}=\sum_{\i\in\It} {w_\i}/{c_{\pa(i)}}.
\end{equation}    
In \eqref{eq:GenExProp} and everywhere else we use the convention that $\prod_{\i\in\emptyset} \ldots=1$. If $\It=\emptyset$ then $\s$ is undefined.

The \textit{extended target} is concentrated on trees with $\It\not=\emptyset$ and is given by
\begin{equation}\label{eq:GenExTarg}
\begin{split} 
 \phi(\I,\E,{\d}\xx,{\d}\tt,\s)&=\psi(\I,\E,{\d}\xx,{\d}\tt,\s)\frac{\hat{z}}{\z}\\
              &=\pi({\d}x_{\an(\s)},{\d}t_{\an(\s)})\cdot \psi_{\rm cond}(\I,\E,{\d}\xx,{\d}\tt,\s|x_{\an(\s)},t_{\an(\s)}),
\end{split}
\end{equation}
where the \textit{conditional proposal distribution} is 
\begin{equation}\label{eq:CondExProp}
\begin{split}
 \psi_{\rm cond}&(\I,\E,{\d}\xx,{\d}\tt,\s|x_{\an(\s)},t_{\an(\s)}) \\                       
                          &=\prod_{\i\in\I\setminus\It\setminus\an(\s)}  \exp\left[-\lambda_{\i} w_{\i}\right]
                                  (\lambda_{\i} w_{\i})^{|\ch(\i)|}
                          \prod\limits_{\j\in\ch(\i)}\ker(x_{\i},t_{\i};{\d}x_{\j},{\d}t_{\j})\\                        
                        &\times 
                          \prod_{\i\in\an(\s)\setminus\{\s\}}  \exp\left[-\lambda_{\i} w_{\i}\right]
                                  (\lambda_{\i} w_{\i})^{|\ch(\i)|-1}
                          \prod\limits_{\j\in\ch(\i)\setminus\an(\s)}\ker(x_{\i},t_{\i};{\d}x_{\j},{\d}t_{\j}).\\      
\end{split}
\end{equation}

Formula \eqref{eq:GenExTarg} plays a crucial role in our paper. 
It relates the result of running PTPF (extended proposal $\psi$) to the extended target $\phi$.
Thus $\phi$ is a probability distribution which, when marginalized to the selected path, yields the target 
distribution $\pi$. It is worth mentioning that \eqref{eq:GenExTarg} is an exact analogue of a fact
established for filters with deterministic number of partices in \citep[see the sentence which follows Theorem 2]{Andrieu2010}. 
Rather unexpectedly, the same relation is true for PTPF.

To verify that equations \eqref{eq:GenExTarg} and \eqref{eq:CondExProp} are correct, it is enough to rearrange terms in $\psi(\I,\E,{\d}\xx,{\d}\tt,\s)\hz/\z$. 
By \eqref{eq:Posterior}, if we gather the terms corresponding to the selected path then we obtain 
\begin{equation}\nonumber
 \prod_{\j\in\an(\s)}\ker(x_{\pa(\j)},t_{\pa(\j)};{\d}x_{\j},{\d}t_{\j})w_{\j}/\z=\pi(\d x_{\an(\s)},\d t_{\an(\s)}).
\end{equation}
Note that the product on the LHS includes $w_{\s}$. Now consider the remaining terms. If $\i\in\an(\s)\setminus\{s\}$ then 
the exponent in the expression  $(\lambda_\i w_\i)^{|\ch(\i)|-1}$ is decreased by one, because one $w_\i$ is included in $\pi(x_{\an(\s)},t_{\an(\s)})$ and 
one $\lambda_\i$ is present in $c_{\pa(\s)}$. In the product of  $\ker(x_{\i},t_{\i};{\d}x_{\j},{\d}t_{\j})$ over the children of $\i$, we drop
one term, which corresponds to $\j\in\an(\s)$, because it is included in $\pi(x_{\an(\s)},t_{\an(\s)})$.  Thus we see that $\psi_{\rm cond}$ in
\eqref{eq:GenExTarg} is indeed given by \eqref{eq:CondExProp}.

Now we can define the conditional PTPF (cPTPF), i.e.\ the algorithm which produces a configuration with the 
probability distribution $\psi_{\rm cond}$. cPTPF differs from the basic PTPF only
in that the conditioning path $(X_{\an(S)},T_{\an(S)})$ is fixed at the beginning and equal to a given
$(X_{1:M},T_{1:M})$.

\begin{center}
 Algorithm cPTPF (conditional PTPF)
\end{center}\smallskip\nobreak
\hrule
\smallskip\nobreak
\algsetup{indent=3em}
\begin{algorithmic}
    \STATE Input $(X_{1:M},T_{1:M})$ 
    \STATE \COMMENT{ \blu{Initialize:} } 
    \STATE $\I:=\Iu:=\{\root\}\cup \{1:M-1\}$;  $\E:=\{k-1\to k, k\in\{1:M\}\}$
    \STATE \COMMENT{ \blu{The values $(X_{1:M},T_{1:M})$ are inherited from the input and kept fixed } } 
    \STATE $T_\root:=t_0$; $C_\root:=\Lambda_\root:=\lambda_0$; $W_\root:=1$ 
    \STATE \COMMENT{ \blu{Main loop:} }
    \STATE $\cdots \cdots \cdots$ \COMMENT{ \blu{ the same as in PTPF} } 
    \STATE $S:=M$   \quad  \COMMENT{ \blu{$S$ identifies the conditioning path } }
    \STATE Output $(\I,\E,\XX,\TT)$ \quad  \COMMENT{ \blu{Tree with the conditional distribution $\psi_{\rm cond}$ } }
    \end{algorithmic}
\nobreak\hrule

A few comments are due here. In the pseudo-code above, we include the conditioning path $(X_{1:M},T_{1:M})$ in the tree with
labels $\{1:M\}$  given to the nodes of this path. Remember that labelling of nodes is arbitrary. 
The only restriction is that in the ``Main loop'', newly created
nodes are given unique labels (different from  $\{1:M\}$). At the last stage of cPTPF, we set ``$S:=M$'' only to ensure
that $(X_{1:M},T_{1:M})=(X_{\an(\S)},T_{\an(\S)})$, in agreement with our notation in \eqref{eq:GenExTarg} and \eqref{eq:CondExProp}.  
\setcounter{section}{3}
\section{Particle MCMC based on PTPF}\label{sec:PMCMC}

The two main Particle MCMC algorithms are \textit{Particle Independent Metropolis-Hastings} and \textit{Particle Gibbs Sampler}.
Their versions with Poisson resampling are algorithms PTMH and PTGS defined below (PT stands for \textit{Poisson Tree}). 
We will describe two recipes for simulating a Markov chain $\Xpp^{(0)},\Xpp^{(1)},\ldots,\Xpp^{(n)},\ldots$, where
$\Xpp^{(n)}=\{\Xpp^{(n)}(\t): \tmin\leq\t\leq\tmax\}$ such that the stationary distribution is the target, i.e.\ 
the posterior of hidden $\Xpp$ given $\Ypp=\yp$. As usual, the trajectories are represented by their skeletons,
so we actually simulate sequences $(\Xp^{(n)},\T^{(n)})=(\Xp^{(n)}_{1:M^{(n)}},\T^{(n)}_{1:M^{(n)}})$, $n=0,1,\ldots$.
The rules of transition from $\Xp =\Xp^{(n)}$ to $\Xp'=\Xp^{(n+1)}$ are the following.
\begin{center}
 One step of PTMH (Poisson Tree Metropolis-Hastings)
\end{center}\smallskip
\hrule
\begin{algorithmic}
    \STATE Input $\hZ,(\Xp_{1:M} ,\T_{1:M})$ \quad \COMMENT{ \blu{Output of the previous step} }
    \STATE Run PFPF to obtain $(\Xp^*_{1:M^*},\T^*_{1:M^*})$ and $\hZ^*$ \quad  \COMMENT{ \blu{Proposal } }
    \STATE Sample $U\sim\uni(0,1)$
    \IF {$U<\hZ^*/\hZ $}
    \STATE $(\Xp'_{1:M'},\T'_{1:M'}):=(\Xp^*_{1:M^*},\T^*_{1:M^*})$; $\hZ':=\hZ^*$\quad \COMMENT{ \blu{Accept} }
    \ELSE 
    \STATE $(\Xp'_{1:M'},\T'_{1:M'}):=(\Xp_{1:M},\T_{1:M})$; $\hZ':=\hZ$ \quad \COMMENT{ \blu{Reject} }
    \ENDIF
    \STATE Output $\hZ'$, $(\Xp'_{1:M'},\T'_{1:M'})$
    \end{algorithmic}
\medskip
\hrule

Our Particle Gibbs Sampler, just as its classical counterpart, can include the additional step of
parent sampling. However, we first describe the basic version (without parent sampling).

\begin{center}
 One step of PTGS (Poisson Tree Gibbs Sampler)
\end{center}\smallskip
\hrule

\smallskip\nobreak

\begin{algorithmic}
    \STATE Input $(\Xp_{1:M},\T_{1:M})$ \quad \COMMENT{ \blu{Output of the previous step} }
    \STATE Run cPFPF to obtain $(\I,\E,\XX,\TT)$ \:  \COMMENT{\blu{Tree with the distribution $\psi_{\rm cond}$}}
    \STATE  \COMMENT{ \blu{Select new $S'$:} }
    \STATE Select $\S'\in\It$ from the probability distribution  $\Pr(\S'=\s')\propto {W_{\s'}}/{\C_{\pa(s')}}$ 
    \STATE Output $(\Xp'_{1:M'},\T'_{1:M'}):=(X_{\an(\S')},T_{\an(\S')})$
    \end{algorithmic}
\nobreak\hrule

In fact, the main results  are straightforward consequences of \eqref{eq:GenExTarg}.

\begin{prop}\label{pr:Z}
Let $f$ be a nonnegative function on the space of  skeletons
$(\xp_{1:m},\t_{1:m})$ and  $\pi (f)=\Ex_{\pi} f(\Xp_{1:M},\T_{1:M})$. If the structure $(\I,\E,\XX,\TT)$ is produced by PTPF then
the following estimator of $\z\pi (f)$ is unbiased:
\begin{equation}\nonumber
 \widehat{\z\pi (f)}=\begin{cases} \displaystyle\sum_{\i\in\It}\dfrac{W_{\i}}{C_{\pa(\i)}}f\left(\Xp_{\an(\i)},\T_{\an(\i)}\right)
                                          & \text{if }\enspace \It\not=\emptyset;\\
                                          & \\
                                        0 & \text{if }\enspace \It=\emptyset.
                   \end{cases} 
\end{equation}
In particular, $\hZ$ is an unbiased estimator of $\z$.
\end{prop}

\begin{proof} By \eqref{eq:GenExTarg}, if $(\I,\E,\XX,\TT,\S)\sim \phi$ then the marginal distribution of 
$(X_{\an(\S)},T_{\an(\S)})$ is $\pi$.
Therefore $\Ex_\phi f(X_{\an(\S)},T_{\an(\S)}))=\pi (f)$. Again using   \eqref{eq:GenExTarg}, we see that
\begin{equation}\nonumber
 \psi(\I,\E,\d\xx,\d\tt)\frac{w_{\s}}{c_{\pa(\s)}}=\hat{z}\psi(\I,\E,\d\xx,\d\tt,\s)=\z\phi(\I,\E,\d\xx,\d\tt,\s),
\end{equation}
where 
$({w_{\s}}/{c_{\pa(\s)}})/\hat{z}=\Pr_\psi(\S=s|\I,\E,\xx,\tt)$. Now it is enough to multiply both sides of the last display
by $f(x_{\an(\s)},t_{\an(\s)})$, integrate over $(\d\xx,\d\tt)$ and sum over $\s\in\It$ to obtain the result. 
Unbiasedness of $\hZ$ follows if we put $f\equiv 1$.
\end{proof}

\begin{thm}\label{th:Main}
Markov chains  generated by algorithms PTMH and PTGS have the equilibrium distribution equal to the target 
$\pi=\pi_{\rm post}$ given by \eqref{eq:Posterior}.  
\end{thm}
\begin{proof}
The line of argument is almost the same as for the classical pMCMC algorithms with multinomial resampling.
The crucial point is equation \eqref{eq:GenExTarg}.

For PTMH, we use equation \eqref{eq:GenExTarg} to infer that
\begin{equation}\nonumber
 \frac{\hat{z}^*}{\hat{z}}=\frac{\phi(\I^*,\E^*,\d\xx^*,\d\tt^*,\s^*)\psi(\I,\E,\d\xx,\d\tt,\s)}{\phi(\I,\E,\d\xx,\d\tt,\s)\psi(\I^*,\E^*,\d\xx^*,\d\tt^*,\s^*)},
\end{equation}
where $\hat{z}^*,\I^*,\E^*,\xx^*,\tt^*,\s^*$ are new values produced by running PTMH, while $\hat{z},\I,\E,\xx,\tt,\s$ are values 
from the previous step. It follows that this algorithm is a proper Metropolis-Hastings procedure 
with the proposal distribution $\psi(\I,\E,\d\xx,\d\tt,\s)$ and the target $\phi(\I,\E,\d\xx,\d\tt,\s)$ on the space of configurations. 
The second equation in \eqref{eq:GenExTarg} shows that the distribution $\phi$ preserved by PTMH has the right marginal distribution 
$\pi(\d x_{\an(\s)},\d t_{\an(\s)})$.

For PTGS, \eqref{eq:GenExTarg} shows that by running cPTPF we sample a configuration with
the conditional distribution $\psi_{\rm cond}(\I,\E,{\d}\xx,{\d}\tt,\s|x_{\an(\s)}=\xp_{1:m},t_{\an(\s)}=\t_{1:m})$. 
If, at the input, $(\Xp_{1:m},\T_{1:m})\sim\pi$ then configuration $(\I,\E,\XX,\TT)$ obtained by
PTPG has the distribution $\phi$ marginalised with respect to $\S$. Consequently, after new $\S'$ has been chosen,
we obtain $(\I,\E,\XX,\TT,\S')\sim \phi$ 
with the marginal $(X_{\an(\S')},T_{\an(\S')})\sim\pi$ at the output. 
\end{proof}
\begin{rem}
 In this section we present particle methods (Particle Metropolis and Particle Gibbs) to sample hidden trajectory given static parameters based on Poisson resampling scheme. 
 Now Bayesian inference on static parameters could be done by the same way as in standard PMCMC methods, for details we refer to \citet{Andrieu2010}.
\end{rem}

\section{Variants and Extensions}\label{sec:Variants}

In this section we present several variants and extensions of the basic algorithms.
In particular we introduce the additional step of ancestor sampling in our particle Gibbs algorithm.
The discussion is focused on two closely related issues.  First is choosing the intensity parameters $\Lambda_\i$. 
Second is parallelisation of computations. 

In our description of  algorithm PTPF, the step of choosing $\Lambda_\i$s was left unspecified. 
We only assumed that $\Lambda_\i=\L(\his(T_\i))$, without any conditions on function $\L$. This
assumption is sufficient to ensure that our algorithms are correct, i.e.\   
the results in Section \ref{sec:PMCMC} and their proofs are valid. However, the efficiency of the 
algorithms crucially depends on the choice of $\Lambda_\i$s.  
The intensity parameters control the size of the population of particles.
It is equally undesirable to allow for an uncontrolled increase and for a rapid decrease 
(or even extinction) of the population. 

One of our objectives is to construct algorithms in which computations are performed in a parallel way.  
In principle, perfectly parallel versions of PTPF, PTMH and PTGS are simple.   
If every parameter $\Lambda_i$ depends only on $\an(\i)$, i.e.\ 
if we set $\L(\his(T_{\i}))=\L(X_{\an(\i)},T_{\an(\i)})$ then
the descendants of $\i$ evolve completely independently of other nodes not belonging to $\an(\i)$. 
However, this scenario is unrealistic, because it makes the number of particles impossible to control.
 
Another scenario is in some sense at the opposite extreme. Suppose that to control the number of particles, we 
allow $\Lambda_i$ to depend on all the particles existing immediately before $T_\i$, i.e.\ 
$\{\j\in \I: T_{\pa(\j)}< T_\i\leq T_{\j}\}$. This makes parallel construction of algorithms much more difficult.   

\subsection{Discrete time models}

We begin with the easier case of discrete time models.
If time is discrete ($t=1,\ldots,m$) then $\I_t=\{\i\in\I: T_\i=t\}$ is $t$th ``generation'' of particles and all $\i\in\I_t$  
propagate simultaneously. It is natural to choose a common value, $\Lambda_\i=\Lambda_t$, for all $\i\in\I_t$.
An obvious way to stablize the number of particles is to choose 
\begin{equation}\label{eq:DiscreteLambdas}
 \Lambda_t=\frac{\lambda_0}{\sum_{\j\in\I_t}W_\j},
\end{equation}
because then  $\Ex |\I_{t+1}|=\sum_{\i\in\I_t} \Lambda_{\i} W_{\i}=\lambda_0$. Our simulations show that the  rule
\eqref{eq:DiscreteLambdas} well stabilizes not only the expected number but also the actual number of particles, see the results 
presented in Section \ref{sec:Sim}. Moreover, PTGS with the rule \eqref{eq:DiscreteLambdas} is uniformly ergodic, under 
the same assumptions as for the standard Particle GS. Theorem \ref{th:UnifErg} below and the method of proof are similar to 
\citep{LindDoucMoul2015}.  We verify a Doeblin condition for one step transition
of discrete time PTGS. {The proof of Theorem \ref{th:UnifErg} is given in the Supplementary Material.} 

Recall that in a single step, PTGS takes a trajectory $X_{1:m}$ and outputs a new trajectory $X'_{1:m}=X_{\an(\S')}$.
The target distribution $\pi=\pi_{\rm prior}$ is given by \eqref{eq:PosteriorDiscr}. Symbol $\Pr$ refers to the the transition probability of PTGS.  

\begin{thm}\label{th:UnifErg} Consider discrete time PTGS with the rule \eqref{eq:DiscreteLambdas}. If the likelihood functions are uniformly bounded, \ie $\Vert \lik_t\Vert_\infty =
\sup_{x_t\in\X}\lik_t(x_t)\leq c<\infty$ then the following minorisation condition holds. For every measurable subset $\D$  of  
$\X^m$ and every $x_{1:m}\in\X^m$ we have
\begin{equation}\nonumber
\Pr(X'_{1:m}\in \D|X_{1:m}=x_{1:m})\geq \eps \pi(\D),
 \end{equation}
for some constant $\eps>0$.
\end{thm}

This theoretical result confirms that under \eqref{eq:DiscreteLambdas}, PTGS is as efficient as its classical counterpart.
Let us remark that Theorem \ref{th:UnifErg} remains valid (with the same proof) also for PTGAS, the version
of PTGS with ancestor sampling to be introduced in the next subsection.

{
Unfortunately, it is difficult to reconcile \eqref{eq:DiscreteLambdas} with the parallel structure
of computations. Some special properties of the Poisson distribution offer a possible way to overcome these difficulties and 
efficiently parallelise computations. Well-known techniques of ``thinning'' and ``superposition'' can be used
in sampling the Poisson tree $\Tree$. We can use some preliminary approximation of 
$\sum_{\j\in\I_t}W_\j$ to compute ``tentative'' value of $\Lambda_t$ at every time $t$. Then, in the next stage,
the tree can be adjusted by sampling additional children and their descendants (superposition) or removing
some children and their descendants (thinning).
Another method is to use only a random sample of existing particles to determine $\Lambda_t$.}

\subsection{Ancestor sampling for discrete time models}

Although algorithm PTGS does preserve $\pi$, its mixing properties are poor because of the well-known phenomenon of path-degeneration (as for the
classical  particle Gibbs Sampler). A remedy is to additionally resample parents, i.e.\
change those arrows in $\E$ which lead to nodes in the (old) selected path. 
We adapt the method proposed in \citep{Lindsten2014} to our Poisson tree setting. 
\goodbreak

For discrete time models, a modification of PTGS is straightforward and as simple as the original ancestor 
sampling in \citep{Lindsten2014}.  The posterior is given by  
\eqref{eq:PosteriorDiscr} and thus $W_\i=\lik_t(X_\i)$ for $\i\in\I_t$. 
Assume that the intensity parameters are given by \eqref{eq:DiscreteLambdas}.  
In the following algorithm PTGAS-dt, we assume that the transition kernels
$\P_{t-1}(\xp_{t-1},\d\xp_{t})$ are represented by transition \textit{densities} $\p_{t-1}(\xp_{t-1},\xp_{t})$.

Recall that according to notations used in cPTPF, $\an(S)=(1,\ldots,\t,\ldots,m)$. Therefore, in the pseudo-code below, 
instructions $\j:=\t$;  $\i:=\t-1$ mean that we pick up an arrow belonging to the conditioning path (the input of cPTPF).
Recall that $\I_t=\{\i\in\I: T_\i=t\}$.
\begin{center}
 One step of PTGAS-dt (Poisson Tree Gibbs with Ancestor Sampling - discrete time) 
\end{center}\smallskip
\hrule
\medskip
\begin{algorithmic}
    \STATE Input $\Xp_{1:m}$ \quad \COMMENT{ \blu{Output of the previous step} }
    \STATE Run cPFPF to obtain $(\I,\E,\XX,\S)$ \:  \COMMENT{ \blu{Tree with the distribution $\psi_{\rm cond}$} }
    \STATE $\E'_0:=\E$
    \FOR {$\t=2,\ldots,m$} 
    \STATE Let $\j:=\t$; $\i:=\t-1$ \: \COMMENT{\blu{ $\i\to\j$ is an arrow in the conditioning path}} 
    \STATE Sample $\i'\in \I_{\t-1}$ with probability $\Pr(\i')\propto  W_{\i'} \p_{\t-1}(X_{\i'},X_\j)$  \:  
              \STATE   \COMMENT{\blu{New parent of $\j$}} 
    \STATE $\E'_\t:=(\E'_{\t-1}\setminus\{\i\to\j\})\cup\{\i'\to \j\}$ \quad  \COMMENT{ \blu{Change arrow} }
    \ENDFOR 
    \STATE  $\E':=\E'_{m}$
    \STATE  \COMMENT{ \blu{Select new $S'$:} }
    \STATE Select $\S'\in\It$ from the probability distribution  $\Pr(\S'=\s')\propto {W_{\s'}}$ 
    \STATE Output $\Xp'_{1:m}:=X_{\an'(\S')}$
           \STATE \COMMENT{ \blu{$\an'(\cdot)$ is the ancestor line corresponding to $\E'$, the new set of arrows} }
    \end{algorithmic}
\nobreak\hrule

\begin{thm}\label{th:ASdiscrete}
Markov chain  generated by algorithm PTGAS-dt has the equilibrium distribution equal to the target $\pi$.  
\end{thm}
{The proof is in the Supplementary Material.}

\subsection{Continuous time models}

\def\bmin{\beta_{\rm min}}
\def\bmax{\beta_{\rm max}}
\def\be{\beta}
\def\lam{l}

Choosing the intensity parameters  is more difficult in the case of continuous time models. We have 
$W_\i=\lik(X_{\i};T_{\pa(\i)},T_\i)$, thus $W_\i$ depends on the sample path $\Xpp_i$ in
the time interval $[T_{\pa(\i)},T_\i[$. This means that the weights are actually assigned to arrows,
not to nodes. It is not reasonable to compare likelihoods which correspond to different time intervals,
so a formula analogous to \eqref{eq:DiscreteLambdas} would make little sense. The solution we propose
is in a sense a compromise between the two ``extreme'' scenarios sketched in the first part of this section.
Roughly speaking, we partition the interval $[\tmin,\tmax]$ into subintervals or ``strips''. The particles within every 
strip evolve independently. At the end of the strip we synchronise the particles 
and compute some statistic which is used to determine $\Lambda_\i$s in the next strip.

We proceed to details. The points of partition (arbitrarily chosen) are 
\begin{equation}\nonumber
 \tmin=\ts^0<\ts^1<\cdots<\ts^r<\cdots<\ts^q=\tmax
\end{equation}
($\ts$ standing for  `synchronisation time'). Let 
\begin{equation}\nonumber
 \F^r=\{\i: \ts^r\leq T_\i<\ts^{r+1}\}. 
\end{equation}
If $\i\in\F^r$ then we say that particle $\i$ is in $r$th strip, i.e.\ has a chance to propagate in the interval 
$[\ts^r,\ts^{r+1}[$. Let 
\begin{equation}\nonumber
 \F^r_\circ=\{\i: T_{\pa(\i)}<\ts^r\leq T_\i<\ts^{r+1}\} \text{ and } \G^r=\{\i: T_{\pa(\i)}<\ts^r, T_\i\geq\ts^{r+1}\}. 
\end{equation}
Note that $\F^r\setminus \F^r_\circ$ is the set of nodes in $\F^r$ whose parents are also in $\F^r$. 
The number of particles that exist immediately before time $\ts^r$ is $|\F^r_\circ\cup\G^r|$. For every $\i\in\F^r_\circ$, let
\begin{equation}\nonumber
 W_\i^r=\lik(\Xpp_{\i[\ts^{r-1},\ts^r[})
\end{equation}
be the partial likelihood corresponding to the path $\Xpp_{\i}$ in the \textit{previous} strip (whilst $W_\i=\lik(\Xpp_{\i[\t_{\pa(\i)},t_\i[})$). Let
\begin{equation}\nonumber
 W_\circ^r=\sum_{\i\in\F^r_\circ} W_\i^r.
\end{equation}
Let us emphasise that the likelihoods for different paths are computed for the same time interval $[\ts^{r-1},\ts^r[$.  
Now, we propose the following rule of computing $\Lambda_\i$s in $r$th strip. Choose 
a nondecreasing function $b:\:]-\infty,\infty[\to[0,\infty]$. Assume that $\F^r_\circ\not=\emptyset$ and put
\begin{equation}\label{eq:ContinuousLambdas}
 \Lambda_\i=\begin{cases}
             \dfrac{1}{W_\i}\dfrac{W_\i^r}{W_\circ^r}b\left(\lambda_0-|\G^r|\right) & \text{ for } \i\in\F^r_\circ;\\
              &\\ 
             \dfrac{1}{W_\i}& \text{ for } \i\in\F^r\setminus\F^r_\circ.\\
            \end{cases}
\end{equation}
The idea behind this seemingly complicated formula is simple. To begin with, $\lambda_0$ is the expected \textit{initial} 
number of particles. We would like to keep the number of particles as close to $\lambda_0$ as possible, in the course of
building the tree. To this end, we try to control the Poisson intensities $\Lambda_\i W_\i$. 
Note that under \eqref{eq:ContinuousLambdas} we obtain
\begin{equation}\nonumber
 \sum_{\i\in\F^r_\circ}\Lambda_\i W_\i=b\left(\lambda_0-|\G^r|\right). 
\end{equation}
This expression is the expected number of children of nodes in $\F^r_\circ$ (conditioned on the history of the process before $\ts^r$).
The second line in \eqref{eq:ContinuousLambdas} implies that for every node in $\F^r\setminus\F^r_\circ$, the expected number of children is one.
Particles corresponding to $\G^r$ ``pass through'' the strip $[\ts^r,\ts^{r+1}[$ unchanged. Putting this together, we see that
the (conditional) expected number of particles  that exist immediately before $\ts^{r+1}$ is 
\begin{equation}\nonumber
 b\left(\lambda_0-|\G^r|\right)+|\G^r|.
\end{equation}
If we put $b(\lam)=\max(\lam,0)$ then the expected number of particles immediately before $\ts^{r+1}$ 
would be equal to $\max(\lambda_0,|\G^r|)$. 
However, if $|\G^r| \geq \lambda_0$ then the particles in $\F^r$ would have zero chance to propagate.
Therefore, a reasonable strategy is to choose e.g.\ $b(\lam)=\max(\lam,b_0)$ for some small constant $b_0>0$.

Equation \eqref{eq:ContinuousLambdas} involves, apart from the quantities specific to node $\i$, only sets $\F^r_\circ$, $\G^r$ and the sum of weights $W_\circ^r$.
This means that we have to identify all the particles which exist at moment $\ts^r$, know their lifespans and locations before we proceed
to processing nodes in $\F^r$. On the other hand, \textit{we need not sort $T_\i$s for $\i\in\F^r_\circ$}. This fact is important for efficient implementation
of the algorithm. Once we create children of nodes in  $\F^r_\circ$, we compute their lifespans and identify nodes in $\F^r\setminus \F^r_\circ$.
Descendants of every node in $\F^r_\circ$ evolve independently until $\ts^{r+1}$, the next moment of synchronisation. 

\subsection{Ancestor sampling for continuous time models}

\def\pro{'}
\def\den{k}

For continuous time models, ancestor sampling is more complicated. 
The general idea is the same as in \citep{Lindsten2014}. In PTGS, we change the parents of nodes along the 
input path in such a way that preserves $\phi$, the extended target distribution. However, the details are
more difficult, due to the complicated rules for computing $\Lambda_i$s. 
Assume that the intensity parameters are given by \eqref{eq:ContinuousLambdas}.  
Our procedure of ancestor sampling is based on the following idea. We change an existing arrow $\i\to \j$ to a new arrow $\i\pro\to \j$ only
if $\j\in\F^r\cap\G^{r-1}\cap\an(\S)$ and furthermore if $\i$ and $\i\pro$ belong to the same ``synchronisation strip'' (say $\F^p$ with $p<r-1$).
Under these conditions we are able to compute the ratio between $\phi$ distributions for the old configuration and the new one. 
In the pseudo-code below we write $W_{\i\pro\to\j}=\lik(X_\j;T_{\i\pro},T_\j)$. 
Assume also that the transition kernel
$\ker(\xp_{k-1},\t_{k-1},\d\xp_{k},\d\t_{k})$ is represented by transition {density} $\den(\xp_{k-1},\t_{k-1},\xp_{k},\t_{k})$.

\begin{center}
 One step of PTGAS-ct (Poisson Tree Gibbs with Ancestor Sampling - continuous time) 
\end{center}\smallskip
\hrule
\medskip
\begin{algorithmic}
    \STATE Input $(\Xp_{1:M},\T_{1:M})$ \quad \COMMENT{ \blu{Output of the previous step} }
    \STATE Run cPFPF to obtain $(\I,\E,\XX,\TT,\S=M)$  \quad  \COMMENT{ \blu{A tree with the conditional distribution $\psi_{\rm cond}$ } }
    \STATE $\E'_0:=\E$
    \FOR {$\t=2,\ldots,M$} 
    \STATE Let $\j:=\t$; $\i:=\t-1$  \: \COMMENT{\blu{$\i\to\j$ is $\t$th arrow along the conditioning path}} 
    \STATE Find $r$ such that $\j\in\F^r$
    \STATE \COMMENT{ \blu{If $\j\not\in\G^{r-1}$ then do nothing} }
    \IF {$\j\in\G^{r-1}$} 
    \STATE Find $p$ such that $\i\in\F^p$ 
    \STATE Sample $\i\pro\in \F^{p}$ with probability  
        \begin{equation}\nonumber
         \Pr(\i\pro)\propto \frac{W_{\i\pro} W_{\i\pro\to\j} \cdot \den(X_{\i\pro},T_{\i\pro};X_\j,T_\j)}{\C_{\pa(\i\pro)}}
        \end{equation}   
    \STATE $\E'_{\t}:=(\E'_{\t-1}\setminus\{\i\to\j\})\cup\{\i\pro\to \j\}$ 
             \STATE  \COMMENT{ \blu{Sample a new parent of $\j$ and change arrows} }
    \ENDIF
    \STATE $\E':=\E'_{M}$
    \ENDFOR 
    \FORALL {$\i\in\It$}
    \STATE Recompute $\C_{\pa'(\i)}=\prod_{\j\in\an'(\i)\setminus\{\i\}}\Lambda_\j$
     \STATE \COMMENT{ \blu{$\an'(\cdot)$ and $\pa'(\cdot)$  correspond to $\E'$, the new set of arrows} }
    \ENDFOR 
    \STATE  \COMMENT{ \blu{Select new $S'$:} }
    \STATE Select $\S'\in\It$ from the probability distribution  $\Pr(\S'=\s')\propto {W_{\s'}}/\C_{\pa'(\s')}$ 
    \STATE Output $(\Xp'_{1:M'},\T'_{1:M'}):=(X_{\an'(\S')},T_{\an'(\S')})$ 
    \end{algorithmic}
\nobreak\hrule

The following theorem shows that the ancestor sampling step is correct.  

\begin{thm}\label{th:AScont}
Markov chain  generated by algorithm PTGAS-ct has the equilibrium distribution equal to the target $\pi$.  
\end{thm}

The proof is based on the following observations:
\begin{itemize}
\item For any $q$ sets of nodes $\F^{q}$, $\G^{q}$ and  $\F^{q}_\circ$ remain unchanged if $\i\to\j$ is replaced by $\i\pro\to\j$
      (because $\i$ and $\i\pro$ belong to the same strip).
\item Since $\j \in \G^{r-1}$ we ensure that $W_{\j}^{r}=\lik(\Xpp_{\j[\ts^{r-1},\ts^r[})$ remains unchanged. 
Indeed, the assumptions in Section \ref{sec:SSM} imply
that $\Xpp_{\j[\t,\T_\j[}$ is uniquely determined by $X_j$ and does
not depend on $\pa(\j)$, provided that $T_{\pa(\j)}< \t$ (see \eqref{eq:EndPoint} and comments following this equation). 
\end{itemize}
{The full proof of Theorem \ref{th:AScont} is relegated to the Supplementary Material. }

\section{Simulation results}\label{sec:Sim}
In this section we examine PTPF's properties in a series of numerical evaluations. To assess the overall correctness of PTPF scheme in discrete time setting 
we compare our implementation with PGAS from \citep{Lindsten2014}. In subsequent sections we investigate properties of continuous time 
PTPF and implementation details.

\subsection{Discrete time models}\label{subsec:DiscreteSim}

Here we study differences between PTPG and classical Gibbs sampler in the setting of  discrete time processes. To this end we examine two state space models --
stochastic volatility model and simple non-linear model considered in \citep{Andrieu2010}. 
Both algorithms have been run on the same sets of starting and observed trajectories.

\subsubsection{A non-linear state space model}\label{non-linearModel}
\begin{figure}[ht]
\centering
\includegraphics[width=.9\textwidth]{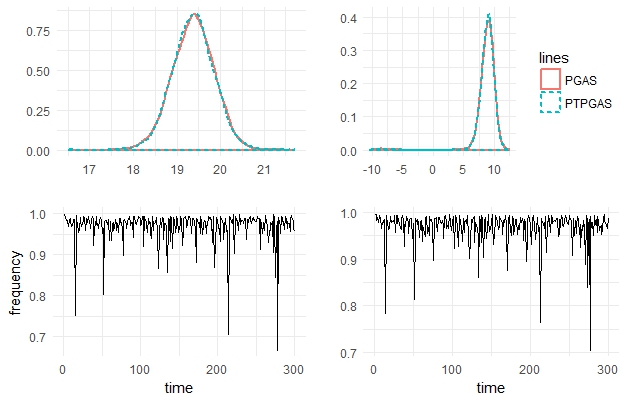}
\caption{Estimated densities of 100'th (upper left corner) and 200'th state (upper right corner). Bottom row - comparison of update frequency for last 1000 iterations between PGAS (left) and PTPGAS (right)}
\label{fig:SSM1}
\end{figure}

\begin{figure}[ht]
\centering
\includegraphics[width=.9\textwidth]{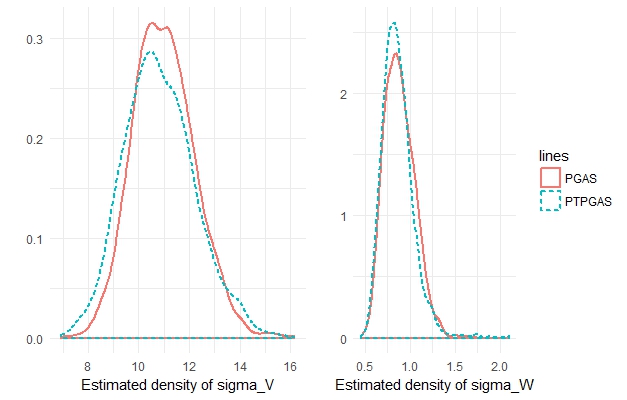}
\caption{Estimated densities of models' parameters in non-liner state space model}
\label{fig:SSM2}
\end{figure}

We start our study with a simple non-linear state space model given by equations:
$$ \Xp_k = \frac{\Xp_{k-1}}{2}  + 25 \frac{\Xp_{k-1}}{1+\Xp_{k-1}^2} + 8\cos(1.2k) + V_k$$
 $$ \Yp_k = \frac{\Xp_{k}^2}{20} + W_k,$$

where $\Xp_1 \sim \mathcal{N}(0,5)$, $V_n \sim \mathcal{N}(0, \sigma_{V}^2)$, $W_n \sim \mathcal{N}(0, \sigma_{W}^2)$.

Priors for both parameters have been set to $\mathcal{IG}(0.01,0.01)$ with $(N,\lambda_0,\T) = (300,300,300)$.
Both algorithms have been run for 10 000 iterations with 3000 burnin and starting parameters equal to: $\sigma^2_V = 10$,  $\sigma^2_W = 1$.

\subsubsection{Stochastic Volatility Model with Leverage}

\begin{figure}[ht]
\centering
\includegraphics[width=.9\textwidth]{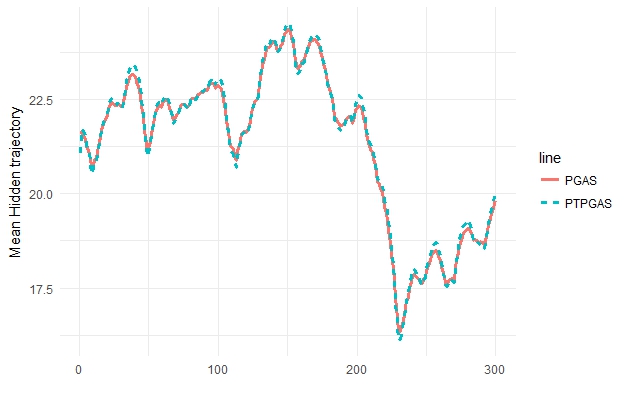}
\caption{Comparison of mean sampled trajectory after burnin of 5000 for stochastic volatility model}
\label{fig:SV1}
\end{figure}

\begin{figure}[ht]
\centering
\includegraphics[width=.9\textwidth]{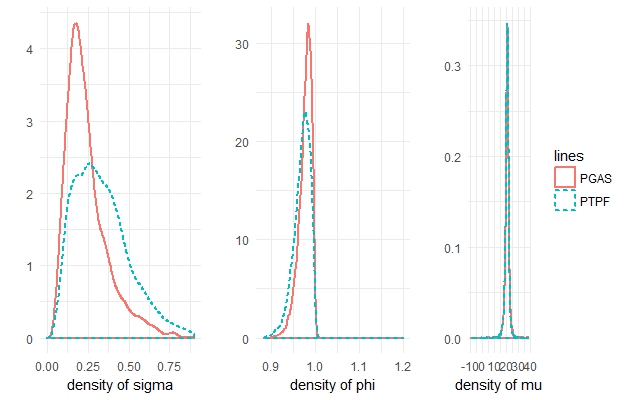}
\caption{Comparison of estimated parameters' densities for stochastic volatility model with leverage}
\label{fig:SV2}
\end{figure}

Next we consider the model governed by equations:
\[\Xp_{k+1} = \mu(1 - \phi) + \phi \cdot \Xp_k + \sigma \cdot \mathcal{N}(0,1)\]
\[\Yp_k = e^{(-0.5\Xp_k)} \cdot \mathcal{N}(0,1)\]

Priors and sampling method were taken from \citep{StochVol}.
Observations have been taken from  Standard and Poor’s (SP) 500 data for the  interval 2017-03-10 -- 2018-05-17.

Both algorithms have been run for 10 000 iterations with 5000 burnin and $(N, \lambda_0) = (1000, 1000)$. Trajectory length has been set to 300.

For both the models we have not found any significant differences between classical particle Gibbs Sampler and 
the Poisson Tree scheme. {Posterior estimates obtained with PTPG may manifest higher share of outliers }-- 
a phenomenon which exhibits itself in 
estimate of $\sigma$ in figure \ref{fig:SV2} -- nonetheless the speed and quality of convergence seems to be comparable. 
Both ancestor sampling schemes seem to provide equivalent improvement in mixing. 
The obvious disadvantage of PTPG is a little bit more involved implementation.

\subsection{Continuous time models}
Here we apply PTPG to two PDSMPs which have been considered among others in \citep{Finkle} 
to illustrate its properties and assess the overall utility of ancestor sampling step.

In both the examples $b$ function (which controls the size of population) has been defined by
\[
b(x)=
\begin{cases}
x, & x \ge 1; \\
0.9x+0.1, & 0 \le x<1; \\
0.1, & x<0.
\end{cases}
\]

\subsubsection{Elementary change-point model}

\begin{figure}[ht]
\centering
\includegraphics[width=.9\textwidth]{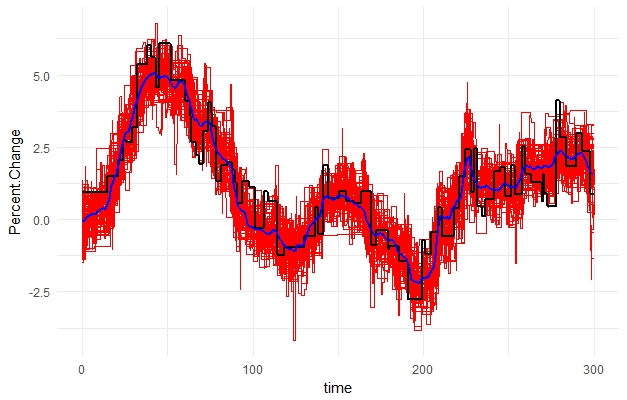}
\caption{Mean sampled trajectory (blue) vs true hidden trajectory (black) for elementary change point model}
\label{fig:Elementary1}
\end{figure}

\begin{figure}[ht]
\centering
\includegraphics[width=.9\textwidth]{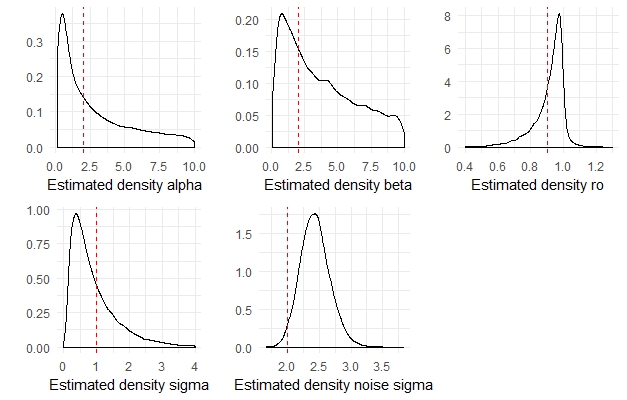}
\caption{Estimated densities of static parameters in elementary change point model}
\label{fig:Elementary2}
\end{figure}

For a first example we have used a simple PDP in which skeleton of $\Xpp$ is assumed to be $AR(1)$ process  
(with coefficient  $\rho$) with unknown variance of noise, $\sigma$.
Jump times are sampled from gamma distribution with unknown parameters $\alpha, \beta$ and mean $\alpha \cdot \beta$. 
Observations are assumed to be taken at ends of fixed time intervals (in our example $[n, n+1[ $ for $n \in \mathbb{N}$) and formed by adding mean $0$ Gaussian noise with~variance~$\sigma_y$~to~$\Xpp$.

Data used for simulations have been sampled with static parameters equal to $(\alpha, \beta, \rho, \sigma, \sigma_y) =(2, 2, 0.9, 1, 2) $.
$\alpha $ and $\beta $ were both given uniform priors on $[0,10]$ whilst the remaining parameters ($ \rho, \sigma, \sigma_y$)  $\mathcal{N}(0,10)$ truncated to $\mathbb{R}_{+}$. 
Synchronization strips were taken to be $[n, n+1[$ for $n \in \mathbb{N}$.

Figure \ref{fig:Elementary1} shows comparison of the mean sampled trajectory with the real hidden trajectory for the last 30 000 iterations of run of 
80 000 iterations (with $\lambda_0 = 3000$). 
For additional reference every $200$th sampled trajectory has been plotted (red colour). 
Posterior distribution of gamma parameters has been sampled with 15000 iterations of independent Metropolis - Hastings algorithm with $Unif(0,10)$ kernel. Remaining posteriors have been approximated with 15000 runs of 
Gaussian random walk Metropolis - Hastings (one run for pair $(\rho, \sigma)$ and one for $\sigma_y$)  with variance of kernel set to 1. 
Estimated posterior densities can be seen in Figure \ref{fig:Elementary2}.

\subsubsection{Shot-noise-Cox-process model}
\begin{figure}[ht]
\centering
\includegraphics[width=.9\textwidth]{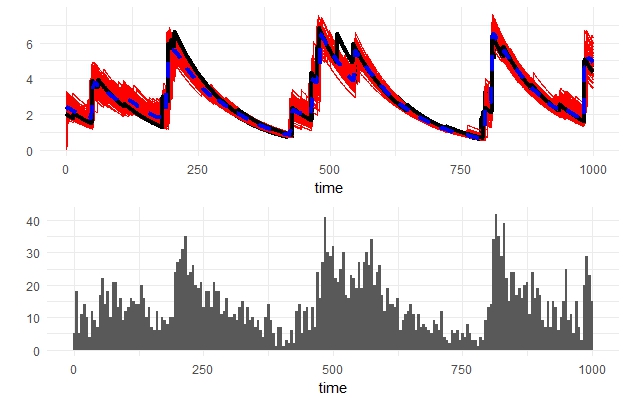}
\caption{Mean (blue) sampled trajectory vs hidden trajectory for Shot-noise-Cox-process model}
\label{fig:MeanVSReal}
\end{figure}

\begin{figure}[ht]
\centering
\includegraphics[width=.9\textwidth]{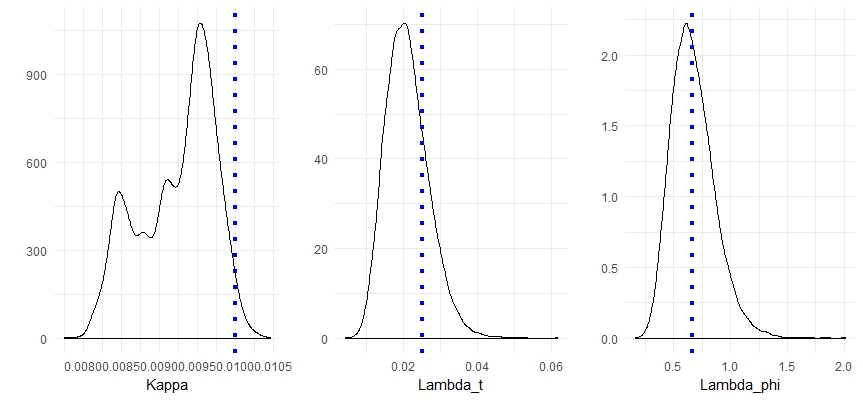}
\caption{Estimated densities of static parameters for Shot-noise-Cox-process model}
\label{fig:Densities}
\end{figure}
The model assumes observations $ \Ypp$ to be an inhomogenous Poisson process with intensity modeled by latent intensity 
$\left( \xpp(t),t \in [0,T] \right)$ sampled from $\Xpp$.

$\Xpp$ is assumed to be a piecewise deterministic process, governed by kernel with density:

\[ \den(x_{n-1}, t_{n-1};  x_{n}, t_n) = \lambda_t \cdot e^{ - \lambda_t\cdot (t_n - t_{n-1})} \cdot\ \mathds{1}_{(t_n - t_{n-1} > 0)} \cdot  \lambda_{\phi} \cdot e^{-\lambda_{\phi} \cdot (x_n -\overline{x}_{n-1} )} \cdot  \mathds{1}_{(x_n -\overline{x}_{n-1} > 0)}\]

where $\overline{x}_{m} = x_m \cdot  e^{-\kappa \cdot (t_m - t_{m-1})}$

For our simulations we have chosen \[\kappa = 0.01 \ , \lambda_t = \frac{1}{40} \  ,\lambda_{\phi} = \frac{2}{3} \  ,T =1000\]
with synchronisation at integer time-points and priors (truncated to $\mathbb{R}_+$)  \[(\kappa, \lambda_t, \lambda_\phi) \sim \mathcal{N}(0,1)\times \mathcal{N}(0,10)\times \mathcal{N}(0,10).\]

Sampling from the posterior distribution has been approximated by 2 runs of Gaussian random-walk Metropolis-Hastings algorithm ($2000$ iterations each) targeting:
\[ \lambda_t \sim \pi(\lambda_t | \xpp, \Ypp) \ (\kappa, \lambda_\phi) \sim \pi(\kappa, \lambda_\phi| \xpp, \Ypp).  \]
(where $\xpp$ is a trajectory sampled in antecedent run of PTPG)

Figures \ref{fig:MeanVSReal}, \ref{fig:Densities} show the result of 80 000 iterations of PTGAS with 50 000 burning and $\lambda_0 = 1000 $.

\begin{figure}[ht]
\centering
\includegraphics[width=.9\textwidth]{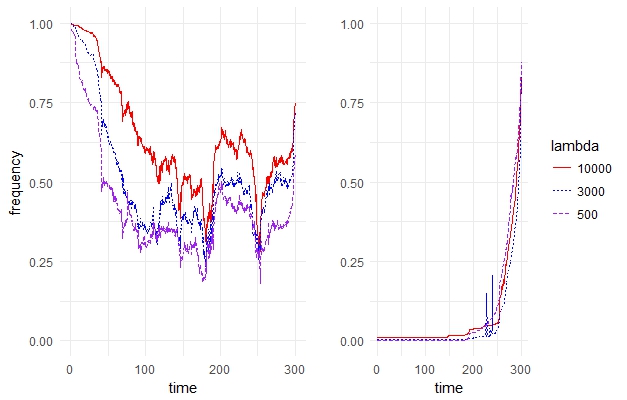}
\caption{Update frequency for 1000 iterations of elementary change point model}
\label{fig:AncestorCont}
\end{figure}

Figure \ref{fig:AncestorCont} depicts update frequency (calculated every $0.5$ time-step) for Poisson Tree Gibbs sampler with (left) 
and without ancestor sampling (right). It is evident that the ancestor sampling step enhances mixing, though the results still look worse than in the discrete time setting. 
Adjustement of synchronisation strips size to jumps' distribution is of crucial importance for ancestor sampling performance.
At least one pilot run is needed to better adjust parameters to data.
\begin{figure}[ht]
\centering
\includegraphics[width=.9\textwidth]{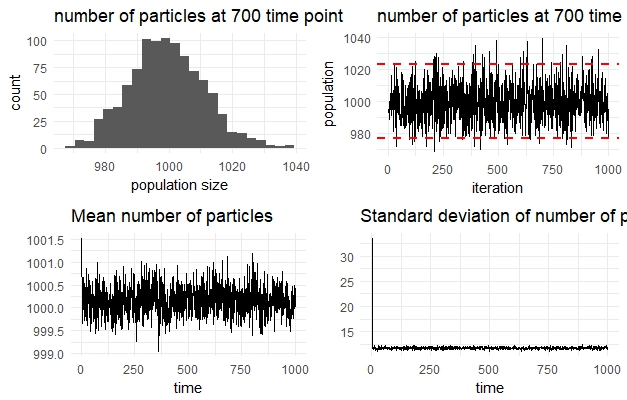}
\caption{Number of particles for 100 iterations}
\label{fig:PopStats}
\end{figure}

We have found that without ancestor sampling the algorithm tends to get stuck at short trajectories (in the sense of number of jumps) for a few iterations. 
This phenomenon is presumably caused by the fact that short trajectories have potentially smaller accumulated ancestors' weights and thus their final weights  
(i.e. weights used to choose a new fixed trajectory) have an order of magnitude  bigger than that of longer trajectories. 
This may probably be additionally countered by introduction of virtual jumps or adaptive size of synchronisation strips 
but we have not pursued those approaches any further. 

Figure \ref{fig:PopStats} shows data on the number of surviving particles for first 1000 iterations of Shot-noise-Cox-process model.
Top row displays histogram and time series of population size for $700$ time step. $95\%$ of iterations stay withing range of $17$ from desirable magnitude. 
Bottom row depicts mean and standard deviation of number of surviving particles at the end of every synchronisation strip. 
It is clear that the employed strategy is effective at controlling size of population --
after a mild burst at the end of first synchronisation strip number of particles stabilizes barely overshooting $[990,1010]$ interval.

\subsection{Implementation and performance comparison}
\begin{figure}[ht]
\centering
\includegraphics[width=.9\textwidth]{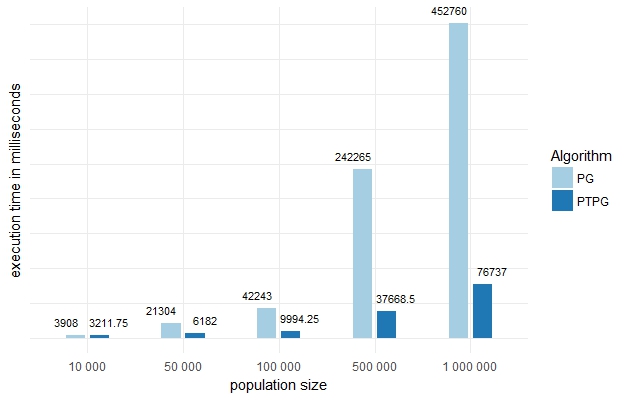}
\caption{Execution time in millisecondsfor PTPG and PG algorithms}
\label{fig:Performance}
\end{figure}

A non-linear state space model from section \ref{subsec:DiscreteSim} has been used for comparison between classical 
Particle Gibbs sampler and Poisson Tree scheme.  The two algorithms have been implemented in C++  and employed utilities from standard library. 
Time measurements have been obtained from runs on 64 cores, 2.5GHz per each.
For both approaches computations were performed by fixed number of threads working in parallel (plus one additional 
thread in case of PTPF whose sole purpose of existence was to gather sums of weights from each thread, 
combine them and redistribute result amongst workers).

Distribution of work between finite number of worker threads is straightforward in the setting of classical Particle Filter 
but gets more troublesome with introduction of PTPF -- one can still divide first population uniformly between threads 
letting each thread take care of its own batch but we have found that after average batch to thread ratio decreases beyond 
a certain point (we have  empirically observed this threshold to be around 500 particles per thread) some batches perish 
completely after few dozens of propagations. To counter this erratic behaviour every 50 steps 
we synchronize all threads and let one chosen thread redistribute surviving particles uniformly. 
Despite the additional overhead introduced by this operation 
we have found this modified scheme to be more effective. The detailed comparisson of our parallel implementation with approaches presented in 
\citep{Paige2014,Murray2016} is disscused in 
Supplementary Material.

Figure \ref{fig:Performance} shows the comparison of mean times of execution (in milliseconds) for 10 iterations, 
trajectory length equal to 400 and number of threads working in parallel equal to 50. 
For big population to thread ratio PTPG provides much faster time of execution. 
However, we have found that its time of execution exhibits larger variance between different runs. 
Both algorithms scale quite well with increasing number of threads but  ordinary Particle Gibbs sampler seems to be more 
resistant to decrease in number of threads -- for instance after reduction to only 10 worker threads execution time for 
one million is around 50s for PG and 20s for PTPG. When number of particles is smaller than 3000 parallel PTPF 
implementation ceases to be practicable.

Implementation of the algorithm for continuous time is considerably more involved, hence an efficient implementation 
encompassing both schemes is not feasible. 
Apart from death time and weight one has to keep information about accumulated parent weight, weight truncated to synchronisation 
strip, time of birth and strip of death.
Additional bookkeeping is neccessary -- for every synchronisation strip $|\G^r|, b(\lambda_0 - |\G^{r}|)$ must be recorded.
Particles which have not propagated are kept on stacks -- one for every synchronisation strip. 
Particles after propagation are stored in vectors (analogously one for every synchronisation strip). 
This enforces much greater movement of data from one place to another (which is essentially constant in discrete time setting).

\section*{Acknowledgement}
The paper is partailly supported by Polish National Science Center grant: NCN UMO-2018/31/B/ST1/00253.

\appendix

\section{Semi-Markov Piece-wise Deterministic Processes} 

\def\z{\mathit{z}}
\def\hz{{\hat{z}}}
\def\hZ{{\hat{Z}}}
\def\Tree{\mathbb{A}}
\def\his{\mathcal{H}}
\def\XX{\mathbf{X}}
\def\WW{\mathbf{W}}
\def\xx{\mathbf{x}}
\def\ww{\mathbf{w}}
\def\TT{\mathbf{T}}
\def\tt{\mathbf{t}}
\def\i{\mathit{i}}
\def\j{\mathit{j}}
\def\l{\mathit{l}}
\def\k{\mathit{k}}
\def\S{\mathit{S}}
\def\s{\mathit{s}}
\def\C{\mathit{C}}
\def\c{\mathit{c}}

\def\bmin{\beta_{\rm min}}
\def\bmax{\beta_{\rm max}}
\def\be{\beta}
\def\lam{l}
\def\tker{\tilde\ker}
\def\tnu{\tilde\nu}
\def\txp{\tilde\xp}
\def\tXp{\tilde\Xp}

Here we provide a more detailed and explicit description of a class of processes we consider.
Let  $\X$ be a Polish space (a complete and separable metric space equipped with its Borel $\sigma$-field).
A \textit{piece-wise deterministic semi-Markov process} (PDSMP) $\Xpp=\{\Xpp(\t),\t\geq\tmin\}$ is a   
process with values in $\X$ which evolves deterministically in continuous time except for a countable collection of stopping times 
at which it randomly jumps. Trajectories of $\Xpp$ are c\`adl\`ag (right continuous functions having left limits). 
Jumps are described by a space-time stochastic transition kernel $\tker=\tker(\xp_{k-1},\t_{k-1};\d \txp_{k-1},\d \t_k)$
and by an initial distribution $\tnu=\tnu(\d\txp_0,\d \t_1)$.
Deterministic dynamics between jumps is described by a function $\det: \X\times [\tmin,\infty[^2\to \X$  which satisfies 
$\det(\xp,\t,\t)=\xp$ and $\det(\xp,\t,\t'')=\det(\det(\xp,\t,\t'),\t',\t'')$ for $\t\leq \t'\leq \t''$. We additionally require
that the function $\t'\mapsto \det(\xp,\t,\t')$ is continuous for all $\xp\in\X$, $\t\in[\tmin,\infty[$ and the map 
$\xp\mapsto \det(\xp,\t,\t')$ is one-to-one for any $\t\leq \t'$. 

Evolution of the process is described in the following steps. The times of jumps are denoted by $\T_1<\cdots<\T_k<\cdots$.
By convention, $\T_0=\tmin$.
Let $\Xp_{k}=\Xpp(\T_{k}-)$ and $\tXp_{k}=\Xpp(\T_{k})$. We define the rules of transitions 
$(\Xp_{k-1},\T_{k-1})\rightarrow (\tXp_{k-1},\T_k) \rightarrow (\Xp_k,\T_{k})$, where the first move is governed by $\tker$
and the second by $\det$: 
\begin{equation}\label{eq:PDSMPskeleton}
\begin{split}
&\Pr(\tXp_{k-1}\in\d \txp_{k-1},\T_{k}\in\d\t_k|\Xp_{k-1}=\xp_{k-1},\T_{k-1}=\t_{k-1})
                                =\tker(\xp_{k-1},\t_{k-1};\d \txp_{k-1},\d \t_k),\\
& \Xp_{k}=\det(\tXp_{k-1},\T_{k-1},\T_k). \\
\end{split}
\end{equation}
Then we put
\begin{equation}\label{eq:PDSMP}
\Xpp(t)=\det(\tXp_{k-1},\T_{k-1},\t), \text{ for }  \T_{k-1}\leq \t< \T_{k}.
\end{equation}
To simplify notation, let us introduce a ficticious state $\xp_0$ and put $\t_0=\tmin$.  
By convention we can write the initial distribution as 
$\tnu(\d\txp_0,\d\t_1)=\tker(\xp_0,\t_0;\d\txp_0,\d\t_1)$. Consequently, equation \eqref{eq:PDSMPskeleton} makes sense
also for $k=1$ and \eqref{eq:PDSMP} completely describes $\Xpp=\{\Xpp(\t),\t\geq\tmin\}$.


In order to facilitate description of our algorithm PTPF and the construction of our ancestor sampling, 
we have chosen to work with the skeleton 
\begin{equation}\label{eq:Skeleton}
  (\Xp_{1}, \T_{1}),\cdots,(\Xp_{k},\T_{k}),\cdots.
\end{equation}
Note that, according to the definitions above, $\Xp_{k}$ is the value \textit{at the end} of $k$th deterministic piece of a trajectory.
The sequence \eqref{eq:Skeleton} is a Markov chain with the transition kernel 
$\ker=\ker(\xp_{k-1},\t_{k-1};\d \xp_{k},\d \t_k)$ implicitly defined via the two transitions in  \eqref{eq:PDSMP}.
Under the convention introduced earlier, the initial distribution can be expressed as
$\Pr(\Xp_1\in\d\xp_1,\T_1\in\d \t_1)=\ker(\xp_0,\t_0;\d\xp_1,\d\t_1)$.
To ensure that \eqref{eq:Skeleton} uniquely defines $\Xpp$, we have assumed that the map 
$\xp\mapsto \det(\xp,\t,\t')$ is one-to-one for any $\t\leq \t'$. Consequently, $\tXp_{k-1}$ is a function
of $(\Xp_{k},\T_{k-1},\T_{k})$. Let us note that in order to recover the trajectory $(\Xpp(t), \T_{k-1}\leq t<\T_k)$,
we have to follow the deterministic dynamics in the reverse direction, starting from $\Xp_k=\Xpp(\T_k-)$ and
proceeding backwards. This is perhaps not easy in general but feasible in many concrete models. For the
important class of piece-wise constant processes it is trivial.

A wide subclass of PDSMPs consists of continuous time \textit{piece-wise deterministic Markov processes}  (PDMPs).
Assume that we have a nonnegative function $\Q$ on $\X\times[\tmin,\infty[$, interpreted as the intensity of jumps and a family
of kernels $\P_{t}=\P_{t}(\xp,\d\tilde{\xp})$ which govern state transitions. The transition rules \eqref{eq:PDSMP} now reduce to
\begin{equation}\nonumber
\begin{split}
&\Pr(\tilde{\Xp}_{k-1}\in\d\txp_{k-1}|\Xp_{k-1}=\xp_{k-1},\T_{k-1}=\t_{k-1})=\P_{\t_{k-1}}(\xp_{k-1},\d\txp_{k-1}),\\
&\Pr(\T_{k}>\t|\tXp_{k-1}=\txp_{k-1},\T_{k-1}=\t_{k-1})=\exp\left[-\int_{\t_{k-1}}^{\t}\Q(\det(\txp_{k-1},\t_{k-1},\u),\u)\d\u\right],\\
&\Xp_{k}=\det(\tilde{\Xp}_{k-1},\T_{k-1},\T_{k}).
\end{split} 
\end{equation}
It is easily seen that the continous time process $\Xpp$ given by \eqref{eq:PDSMP} is Markov (in general, inhomogeneous in time). 
A rigorous proof can be found in \citep{Davis1984}.
Homogeneous PDMPs obtain if the intensity function does not depend on time, i.e.\ $\Q(\xp,\t)=\Q(\xp)$, the kernels $\P_\t$ do not depend on $\t$
 and $\det(\xp,\t,\t')$ depends on $(\t,\t')$ only through $\t'-\t$. In particular, our setup covers \textit{piece-wise constant homogeneous} Markov processes.
In this important special case we have $\det(\xp,\t,\t')=x$, $\tilde{\Xp}_{k-1}=\Xp_{k}$ and
\begin{equation}\nonumber
\begin{split}
&\Pr(\Xp_{k}\in\d\xp_{k}|\Xp_{k-1}=\xp_{k-1},\T_{k-1}=\t_{k-1})=\P(\xp_{k-1},\d\xp_{k}),\\
&\Pr(\T_{k}>\t|\Xp_{k}=\xp_{k},\T_{k-1}=\t_{k-1})=\exp\left[-(\t-\t_{k-1})\Q(\xp_{k})\right].\\
\end{split} 
\end{equation}

Now we proceed to models in which  process $\Xpp$ is hidden and we observe a random element $\Yp$ which depends on $\Xpp$.
The likelihood is the probability of observing $\Ypp=\ypp$, given sample path $\xpp$ of $\Xpp$.
Since $\ypp$ is fixed and need not be explicitly indicated, it will be dropped from notation whenever no misunderstanding can occur.
We are going to describe typical forms of likelihood functions $\lik(\xpp_{[\t,\t'[})$, 
where $\xpp_{[\t,\t'[}=\{\xpp(\t), \t\leq\t< \t'\}$. We always require that these functions satisfy the condition
\eqref{eq:LikBasic}: or $t<t'<t''$,
\begin{equation}\nonumber
 \lik(\xpp_{[\t,\t''[})=\lik(\xpp_{[\t,\t'[})\lik(\xpp_{[\t',\t''[}).
\end{equation}

In many applications, the observation process $\Ypp$ is just a sequence of ``noisy measurements'' of the process $\Xpp$ at
discrete ``observation times'', say $\tmin\leq\tobs^1<\cdots<\tobs^p\leq\tmax$. Formally, we assume that $\Ypp=(\Yp_1,\ldots,\Yp_p)$,
where each $\Yp_r$ is sampled independently from a (possibly time-dependent) probability density $\lik_{r}(\cdot|\xpp(\tobs^r))$.
The likelihood functions in this model are given by
\begin{equation}\label{eq:LikDiscreteObs} 
 \lik(\xpp_{[\t,\t'[})=\prod_{\t\leq\tobs^{r}<\t'}\lik_{r}(\yp_r|\xpp(\tobs^r))
\end{equation}
and clearly fulfil the assumption \eqref{eq:LikBasic}. 
A standard example would be adding a Gaussian noise to observations on a hidden continuous time Markov process, 
see for example simple prey-predator model considered in \citep{LotkaVolterr}. 

Another form of likelihood may be obtained if $\Ypp$ is a continuous time Markov process. Assume that the state space $\Y$ of this process is finite and the transition
intensities of $\Ypp$ depend on a current state of $\Xpp$. Let $Q_\Ypp(v,v'|x)$ be the intensity of transitions from 
$v\in\Y$ to $v'\in\Y$, $v'\not=v$,  if
$\Xpp(t)=x$. The intensity of jumps out of $v$ is $Q_\Ypp(v|x)=\sum_{v'\not=v}Q_\Ypp(v,v'|x)$. 
For definiteness, assume that trajectories of $\Ypp$ are right continuous. 
If we observe $\Ypp=\ypp$ then the likelihood is given by 
\begin{equation}\label{eq:LikCTBN}
  \lik(\xpp_{[\t,\t'[})=\prod_{\substack{\u: \ypp(\u-)\not=\ypp(\u)\\ \t\leq\u\leq\t'}} Q_\Ypp(\ypp(\u-),\ypp(\u)|\xpp(\u))
      \exp\left[-\int_{\t}^{\t'}\Q_\Ypp(\ypp(u)|\xpp(\u))\d\u\right]
\end{equation}
and fulfils the assumption \eqref{eq:LikBasic}.
Equation \eqref{eq:LikCTBN} arises in the context of Continuous Time Bayesian
Networks (CTBNs), c.f.\ \citep{Nod4}. Processes $\Xpp$ and $\Ypp$ may correspond to hidden and observed nodes of 
a CTBN, respectively. Then the posterior distribution of $\Xpp$ describes ``probabilistic inference'' about the behaviour of
the hidden nodes. Monte Carlo methods for CTBNs are subject of articles \citep{Nod1, Nod4, RaoTeh2013a}. Our algorithms
based on Poisson resampling can also be used for CTBNs.

To conclude this section, note that discrete time models can be considered as a special case of continuous time models.
Let $\Xpp=(\Xp_1,\ldots,\Xp_m)$ be a discrete time Markov chain (in general, inhomogeneous in time) with one-step
transition kernels $\P_1,\ldots,P_{m-1}$. Using a convention explained earlier, let us express the 
the initial distribution as $\Pr(\Xp_1\in\d\xp_1)=\P_0(\xp_0,\d\xp_1)$ for a fictitious state $\xp_0$ and 
write $\Pr(\Xp_\t\in \d\xp_\t|\Xp_{\t-1}=\xp_{\t-1})=\P_{\t-1}(\xp_{\t-1},\d\xp_t)$, for $t=1,\ldots,m$.
Of course, $\Xpp$ can be identified with the continuous time process which is equal to  $\Xp_\t$ on the interval
$[\t-1,\t[$. (To keep the notation consistent, put  $\tmin=0$, $\tmax=m-1$, 
$\ker(\xp_{\t-1},\t-1;\d\xp_\t,\{\t\})=\P_{\t-1}(\xp_{\t-1},\d\xp_{\t})$ and $\T_{\t}=\t$ for $t=1,\ldots,m$.) 
The natural assumption about the process of observations in the discrete time  setting is that $\Ypp=(\Yp_1,\ldots,\Yp_m)$, where $\Yp_t$ 
depends only on \textit{one} state $\Xp_{t}$ of the Markov chain.
The likelihood is of the form $\lik_\t(\xp_{t})=\lik_{\t}(\yp_t|\xp_{t})$.


\section{Illustrative Examples}

We provide two simple examples which illustrate the main idea behind our algorithms
and the key elements of the proofs. 

\subsection{Why Poisson resampling works}\label{sec:Why}

The example presented in this subsection basically corresponds to a single ``propagation'' step of 
algorithm PTPF. 
Consider an importance sampling procedure with Poisson resampling.  
Let $\p$ be a probability density on space $\X$ equipped with measure $\d x$. The target density is
\begin{equation}\nonumber
 \pi(x)=\frac{\p(x)w(x)}{\z},
\end{equation}
where $w$ is a weight (importance) function and $\z=\int_\X \p(x)w(x)\d x$. We interpret $\p$ as a prior distribution and
$w$ as the likelihood of observing $y$ given $x$ ($w(x)=\ell(y|x)$, where $y$ is fixed). In this 
interpretation, $\pi$ becomes the posterior distribution. 

The sampling scheme is the following. 
Draw $N\sim  \poi(\lambda)$. If $N=0$ then do nothing and put $\hZ=0$. If
$N>0$ then draw indepedently $X_1,\ldots,X_N\sim \p(\cdot)$. Put $\hZ=\sum_{{j}=1}^N w(X_{j})/\lambda$.
It is obvious that $\Ex\hZ=\z$. 
Choose $S\in\{1:N\}$ with probability
\begin{equation}\nonumber
 \Pr(S=s|N,X_1,\ldots,X_N)=\frac{w(X_s)}{\sum_{{j}=1}^N w(X_{{j}})}.
\end{equation}
We say that the joint probability distribution of all the random variables generated in such a way is
the \textit{extended proposal}. It is denoted by $\psi$ and given by
\begin{equation}\label{eq:ExProp}
 \psi(n,x_1,\ldots,x_n,s)=\e^{-\lambda} \frac{\lambda^n}{n!}\prod_{j=1}^n \p(x_{j})\frac{w(x_s)}{\sum_{j=1}^n w(x_{j})}
\end{equation}
for $n>0$ and $\psi(0)=\e^{-\lambda}$. Note that $\psi$ is defined
on the space $\{0\}\cup\bigcup_{n=1}^\infty \{n\}\times\X^n\times\{1:n\}$.

Define the \textit{extended target} probability distribution $\phi$ by
\begin{equation}\label{eq:ExTarg}
 \phi(n,x_1,\ldots,x_n,s)=\e^{-\lambda} \frac{\lambda^n}{n!}\prod_{j=1}^n \p(x_{j})\frac{w(x_s)}{\lambda \z}
\end{equation}
for $n=1,2,\ldots$ and $\phi(0)=0$. Note that $\phi$ can be decomposed as follows:
\begin{equation}\label{eq:ExTargDecomp}
 \phi(n,x_1,\ldots,x_n,s)=\underbrace{\frac{\p(x_s)w(x_s)}{z}}_{\text{marginal}}\cdot
 \underbrace{\frac{1}{n}\e^{-\lambda} \frac{\lambda^{n-1}}{(n-1)!}\prod_{j\not=s} \p(x_{j})}_{\text{conditional}}.
\end{equation}
\goodbreak

Formula \eqref{eq:ExTargDecomp} shows that $\phi$ is properly normalized and the marginal distribution of $X=X_S$ is exactly $\pi(\cdot)$.
The conditional distribution of all the remaining variables can be obtained in the following way. The number of the other samples,
$N-1$, has the Poisson distribution. Once $N-1$ is selected, we assign $X$ label $S$ chosen uniformly at random from the set $\{1:N\}$, 
then draw $N-1$ samples from $\p(\cdot)$ and assign them labels $\{1:N\}\setminus\{S\}$.
If we start with $X\sim\pi(\cdot)$,  then the conditional sampling scheme  described above produces a configuration 
$(N,X_1,\ldots,X_N,S)$ such that $X=X_S$, and this configuration has the extended target distribution.
From formula \eqref{eq:ExTarg} it is clear that the conditional probability of $S$ given $N,X_1,\ldots,X_N$ is proportional to
$w(X_S)$. If we select new $S'$ from this probability distribution then $X_{S'}\sim\pi(\cdot)$. 
The update $X$ to $X_{S'}$ is just a  single step of the \textit{Particle Gibbs Sampler} (PGS) in our simplified example. 
We have thus verified that PGS preserves the target.

To see that the \textit{Particle Independent Metropolis-Hastings} (PIMH) also preserves the target, it is enough to note that
\begin{equation}\nonumber
 \frac{\phi(n,x_1,\ldots,x_n,s)}{\psi(n,x_1,\ldots,x_n,s)}=\frac{\hz}{\z},\qquad \hz=\sum_{{j}=1}^n w(x_{j})/\lambda.
\end{equation}
In our example, PIMH first generates a proposal $(X^*_{S^*},\hZ^*)$, obtained in a new run of the above-described 
(unconditional) sampling scheme. Then PIMH updates $(X_S,\hZ)$ to $(X^*_{S^*},\hZ^*)$ with
probability $[\hZ^*/\hZ]\land 1$. We have verified that this is a valid Metropolis-Hastings update.

Similar arguments are used (in a much more complicated setting) to show that our main algorithms PTGS and PTMH are correct.
\goodbreak

\subsection{An example of Poisson Tree}\label{sub:Basic}

The example presented here explains the basic relation between the extended target
$\psi$ and extended target $\phi$. It also illustrates notation used in our paper. 

For simplicity, assume that the process $\Xpp$ is \textit{piece-wise constant}.
Consequently, in the tree produced by PTPF, constant value $x_\i$ corresponds to the time interval $[t_{\pa(\i)},t_\i[$.
Consider the tree depicted below. In our example we use a special way of labelling nodes with their
full ancestor paths. The artificial root is $\root$.
\def\tend{\dagger}
\begin{equation}\nonumber
\xymatrix{ 
[\tmin \ar@{.}[dd]                 &  t_1  && t_2  & t_{21}  & \qquad t_{22}  & \tmax]\ar@{.}[ddd]&   >\tmax        \\
                  &  1    &&      &    21   &                &                 &  \blu{221}\\
\root\ar@{=>}[rrr]^{x_2}\ar@{.}[d]\ar[ur]^{x_1} &&& \blu{2}\ar@{=>}[rr]^{x_{22}}\ar[ur]^{x_{21}}\ar[drrrr]_{x_{23}}|!{[uurrr];[drrr]}\hole & 
                                              &\blu{22}\ar@{-}[r]_{x_{222}}\ar@{=>}[urr]^{x_{221}}|!{[uur];[dr]}\hole&
                                                                                \ar[r] &222\\
\text{begin}        &       &&      &         &                 &   \text{end}   & 23\\ 
} 
\end{equation}
Selected terminal node is $221$ and the corresponding sample path is $x_2,x_{22},x_{221}; t_2,t_{22}, t_{221}$ 
(nodes indicated in \blu{blue} and edges with double arrows).

The \textit{extended proposal} (probability of sampling the depicted configuration) in our example is
{\small
\begin{equation}\nonumber
\begin{split}
 \psi(\I,\E,{\d}\xx,{\d}\tt,\s=221)&=\: \exp\left[-\lambda_0\right](\lambda_0)^2\ker(x_\root,t_0;{\d}x_{1},{\d}t_{1})
                               \blu{\ker(x_\root,t_0;{\d}x_{2},{\d}t_{2})}\\
                                            &\times \exp\left[-\lambda_1 w_1\right]\\
                                            &\times \exp\left[-\lambda_2 w_2\right](\lambda_2\blu{w_2})^3
                                                                          \ker(x_2,t_2;{\d}x_{21},{\d}t_{21})
                                                                     \blu{\ker(x_2,t_2;{\d}x_{22},{\d}t_{22})}\\
                                            & \qquad\qquad\qquad\qquad\qquad\qquad\qquad\qquad\qquad\quad \ker(x_2,t_2;{\d}x_{23},{\d}t_{23})\\ 
                                            &\times \exp\left[-\lambda_{21} w_{21}\right]\\  
                                            &\times \exp\left[-\lambda_{22} w_{22}\right](\lambda_{22}\blu{ w_{22}})^2 
                                                                      \blu{\ker(x_{22},t_{22};{\d}x_{221},{\d}t_{221})}
                                                                      \ker(x_{22},t_{22};{\d}x_{222},{\d}x_{222})\\
                                            &\times \frac{1}{\hz}\cdot\frac{\blu{w_{221}}}{\lambda_0\lambda_2\lambda_{22}},\\     
\end{split}
\end{equation}}
where 
{\small
\begin{equation}\nonumber
\hz=\frac{w_{221}}{\lambda_0\lambda_2\lambda_{22}}+\frac{w_{222}}
{\lambda_0\lambda_2\lambda_{22}}+\frac{w_{23}}{\lambda_0\lambda_2}.
\end{equation}}
The \textit{extended target} is
{\small
\begin{equation}\nonumber
\begin{split}
 \phi(\I,\E,{\d}\xx,{\d}\tt,\s=221)&=\psi(\I,\E,{\d}\xx,{\d}\tt,\s=221)\frac{\hz}{\z}\\
                                          &= \blu{\frac{1}{\z} \ker(x_\root,t_0; {\d}x_{2},{\d}t_{2})w_2
                                                \ker(x_2,t_2;{\d}x_{22},{\d}t_{22})w_{22}}
                                          \blu{\ker(x_{22},t_{22};{\d}x_{221},{\d}t_{221}) w_{221}}\\
                                          &\times \exp\left[-\lambda_0\right](\lambda_0)^1\ker(x_\root,t_0;{\d}x_{1},{\d}t_{1})\\
                                          &\times \exp\left[-\lambda_1 w_1\right]\\
                                          &\times \exp\left[-\lambda_2 w_2\right](\lambda_2 w_2)^2
                                                                          \ker(x_2,t_2;{\d}x_{21},{\d}t_{21})
                                                                          \ker(x_2,t_2;{\d}x_{23},{\d}_{23})\\ 
                                           &\times \exp\left[-\lambda_{21} w_{21}\right]\\  
                                           &\times \exp\left[-\lambda_{22} w_{22}\right](\lambda_{22} w_{22})^1
                                                               \ker(x_{22},t_{22};{\d}x_{222},{\d}t_{222})\\                                                       
      &=\blu{\pi({\d}x_{2},{\d}x_{22},{\d}x_{221};\, {\d}t_2,{\d}t_{22},{\d}t_{22},{\d}t_{221})} \\
                              &\times  \psi_{\rm cond}(\I,\E,{\d}\xx,{\d}\tt|x_2,x_{22},x_{221}; t_2,t_{22},t_{221}).             
\end{split}
\end{equation}}
In the above formulae, terms indicated in \blu{blue} correspond to the selected path. The second formula follows from the first one
via rearrangement of blue terms.

The weights are given by    
\begin{itemize}
 \item $w_1=\lik(x_1;t_0,t_1)$, $w_2=\lik(x_2;t_0,t_2)$, 
 \item $w_{21}=\lik(x_{21};t_2,t_{21})$, $w_{22}=\lik(x_{22};t_2,t_{22})$, $w_{23}=\lik(x_{23};t_{2},t_{23})$,
 \item $w_{221}=\lik(x_{221};t_{22},t_{221})$, $w_{222}=\lik(x_{221};t_{22},t_{221})$.
\end{itemize}

To illustrate our definition of the history, consider e.g.\ node 21. We have $\lambda_{21}=\L(\his(t_{21}))$, where
$\his(t_{21})=\{0;\: 1,0\to 1, x_1,t_1;\: 2,0\to 2, x_2,t_2;\: 21,2\to 21, x_{21},t_{21};\: 
              22,2\to 22, x_{22},t_{22};\: 23,2\to 23, x_{23},t_{23}\}$.

Finally note that the equivalence class in our example contains $2\cdot 3\cdot 2$ trees which differ from the depicted one by
different numbering of $\ch(\root)$, $\ch(2)$ and $\ch(22)$. This is why factors $2, 3, 2$ are omitted in the Poisson probabilities
in the formula for $\psi$. 
\goodbreak

\section{Proofs}

We give proofs omitted in our paper.

\subsection*{Proof of Theorem 2}

Define a sequence of nonnegative measurable functions $f_0,f_1,\ldots,f_m$ by backward induction as follows.
Begin with $f_m(x_{1:m})=\Ind(x_{1:m}\in\D)$ and let 
\begin{equation}\nonumber
 f_{t-1}(x_{1:t-1})=\int_\X f_{t}(x_{1:t-1},x_t)\lik_t(x_t)\P_{t-1}(x_{t-1},\d x_t),
\end{equation}
so that $f_{t}$ is a function on $\X^t$. In agreement with our convention introduced in Section \ref{sec:SSM}, the 
above equation extends also to $t=1$ ($f_0$ is really a scalar, being formally ``a function of the fictitious $x_0$'').
Note that 
\begin{equation}\nonumber
 f_0=\z\pi(\D). 
\end{equation}
Define also a sequence of nonnegative measurable functions $h_0,h_1,\ldots,h_m$ by backward induction. Let $h_m(x_m)=1$ and
\begin{equation}\nonumber
 h_{t-1}(x)=\int_\X h_{t}(x_t)\lik_t(x_t)\P_{t-1}(x,\d x_t)+\frac{2}{\lambda_0}\Vert h_t\lik_t\Vert_\infty,
\end{equation}
so that $h_{t}$ is a function on $\X$ and, by convention, $h_0$ is a scalar. 

Recall that $\I_t=\{\j\in\I: T_\j=t\}$ for $t=1,\ldots,m$. The main ingredient of the proof is 
the following inequality.  For $t=1,\ldots,m$,
 \begin{equation}\label{lem:Cak}
 \Ex\dfrac{\sum_{\j\in\I_t} W_\j f_t(X_{\an(\j)})}{\sum_{\j\in\I_t} W_\j h_t(X_{\i})}\geq 
 \Ex\frac{\sum_{\i\in\I_{t-1}} W_\i f_{t-1}(X_{\an(\i)})}{\sum_{\i\in\I_{t-1}} W_\i h_{t-1}(X_{\i})}.
\end{equation}
Note that for $t=1$ the RHS of this inequality reduces to $f_0/h_0$ (we again recall the convention about
the ficticious state at $t=0$, so we can put $\I_0=\{\root\}$ and $W_0=1$).

To prove \eqref{lem:Cak} we first observe that sampling  particles in $\I_t$ can be equivalently
done as follows. 
\def\lam{\lambda_{0}}
\def\jo{{\j_0}}
\begin{itemize}
 \item First we sample $N-1\sim\poi(\lam)$ and create $N-1$ particles in $\I_{t}$ (note that $|\I_t|=N$, since we always
 have one particle in the conditioning path $x_{1:m}$; in accordance with the pseudo-code for cPTGS this particle corresponds to node labelled $\j=t$).
 \item If $N-1=0$ then $\I_t=\{t\}$. Otherwise, for every $\j\in\I_t\setminus\{t\}$ we choose  its parent with probability
 $\Pr(\pa(\j)=\i)\propto W_\i$. Then, of course, sample $X_\j\sim\P_{t-1}(X_\i,\cdot)$.
\end{itemize}
Indeed, every node $\i\in\I_{t-1}\setminus\{t-1\}$ has number of children equal to $N_\i\sim\poi(\Lambda_{t-1}W_\i)$ and 
for $\i=t-1$ we have $N_\i-1\sim\poi(\Lambda_{t-1}W_\i)$, see \eqref{eq:CondExProp}. Our rule \eqref{eq:DiscreteLambdas} entails 
$\Lambda_{t-1}\sum_{\i\in\I_{t-1}}W_\i=\lam$. Consequently, $N-1=\sum_{\i\in\I_{t-1}}N_\i-1\sim\poi(\lam)$, 
and our claim follows from the well-known property of the Poisson distribution. 

We can say that conditionally, given $N$, propagation of $(t-1)$th generation particles 
in cPTPF is identical as in the classical conditional PF with deterministic number of particles
and with multinomial resampling. In particular, new particles of $t$th generation are conditionally independent, 
identically distributed. This fact will be used in the inequalities to follow.

To lighten notation write $\H=\H{(t-1)}$ for (the $\sigma$-field generated by) the history up to $t-1$ and $N=|\I_t|$.
In the formula below we denote by $\jo$ an arbitrarily chosen node in $\I_t\setminus\{t\}$, if $|\I_t|>1$. If
$|\I_t|=N=1$ then the expression involving (unspecified) $\jo$ is equal to $0$. Analogous remark applies to 
$\sum_{\j\in\I_t,\j\not=\jo,\j\not=t}[\cdots]$.
If $N\leq 2$ then the sum is ``empty'' and, by convention, equal to $0$ . In the third line below we use the 
Jensen inequality combined with the fact that the numerator and denominator are independent.
\begin{equation}\nonumber
\begin{split}
 \Ex\left[\frac{\sum_{\j\in\I_t}W_\j f_t(X_{\an(\j)})}{\sum_{\j\in\I_t}W_\j h_t(X_{\j})}\Bigg| \H,N\right]
 &\geq \Ex\left[\frac{(N-1)\;W_\jo f_t(X_{\an(\jo)})}{\sum_{\j\in\I_t}W_\j h_t(X_{\j})}\Bigg| \H,N\right]\\
 &\geq \Ex\left[\frac{(N-1)\;W_\jo f_t(X_{\an(\jo)})}{\sum_{\j\in\I_t,\j\not=\jo,\j\not=t}W_\j h_t(X_{\j})
                                                              +2\Vert\lik_t h_t\Vert_\infty}\Bigg| \H,N\right]\\
 &\geq \frac{(N-1)\;\Ex\left[ W_\jo f_t(X_{\an(\jo)})\Big| \H,N\right]}
   {\Ex\left[\sum_{\j\in\I_t,\j\not=\jo,\j\not=t}W_\j h_t(X_{\j})\Big| \H,N\right] +2\Vert\lik_t h_t\Vert_\infty} \\
 &= \frac{(N-1)\;\Ex\left[ W_\jo f_t(X_{\an(\jo)})\Big| \H,N\right]}
   {(N-2)_+\: \Ex\left[W_\j h_t(X_{\j})\Big| \H,N\right] +2\Vert\lik_t h_t\Vert_\infty}. \\    
\end{split}
\end{equation}
Now note that 
\begin{equation}\nonumber
\begin{split}
 \Ex\left[ W_\jo f_t(X_{\an(\jo)})\Big| \H,N\right]&=
 \sum_{\i\in\I_{t-1}}\frac{W_\i}{W^{(t-1)}}\int\lik_t(x)f_t(X_{\an(\i)},x)\P_{t-1}(X_\i,\d x) \\
 &=\sum_{\i\in\I_{t-1}} \frac{W_\i}{W^{(t-1)}} f_{t-1}(X_{\an(\i)})
\end{split}
\end{equation}
and, analogously, 
\begin{equation}\nonumber
\begin{split}
 \Ex\left[ W_\j h_t(X_{\j})\Big| \H,N\right]&=
 \sum_{\i\in\I_{t-1}}\frac{W_\i}{W^{(t-1)}}\int\lik_t(x)h_t(X_{\i},x)\P_{t-1}(X_\i,\d x) \\
 &=\sum_{\i\in\I_{t-1}} \frac{W_\i}{W^{(t-1)}} \tilde h_{t-1}(X_{\i}),
\end{split}
\end{equation}
where $W^{(t-1)}=\sum_{\i\in\I_{t-1}} W_\i$ and $\tilde h_{t-1}=h_{t-1}-\frac{2}{\lam}\Vert\lik_t h_t\Vert_\infty$. Consequently, 
\begin{equation}\nonumber
\begin{split}
 \Ex\left[\frac{\sum_{\j\in\I_t}W_\j f_t(X_{\an(\j)})}{\sum_{\j\in\I_t}W_\j h_t(X_{\j})}\Bigg| \H,N\right]
 &\geq \frac{(N-1)\;\sum_{\i\in\I_{t-1}} {W_\i} f_{t-1}(X_{\an(\i)})}
   {(N-2)_+\: \sum_{\i\in\I_{t-1}} {W_\i} \tilde h_{t-1}(X_{\i}) +2\sum_{\i\in\I_{t-1}} {W_\i} \Vert\lik_t h_t\Vert_\infty}. \\ 
\end{split}
\end{equation}
Dropping $N$ from the condition we obtain, by Jensen inequality,
\begin{equation}\nonumber
\begin{split}
 &\Ex\left[\frac{\sum_{\j\in\I_t}W_\j f_t(X_{\an(\j)})}{\sum_{\j\in\I_t}W_\j h_t(X_{\j})}\Bigg| \H\right]\\
 &\geq \sum_{n=1}^\infty \e^{-\lam}\frac{\lam^{n-1}}{(n-1)!}\cdot\frac{(n-1)\;\sum_{\i\in\I_{t-1}} {W_\i} f_{t-1}(X_{\an(\i)})}
   {(n-2)_+\: \sum_{\i\in\I_{t-1}} {W_\i} \tilde h_{t-1}(X_{\i}) +2\sum_{\i\in\I_{t-1}} {W_\i} \Vert\lik_t h_t\Vert_\infty} \\
 &=\lam  \sum_{n=2}^\infty \e^{-\lam}\frac{\lam^{n-2}}{(n-2)!} \cdot\frac{\sum_{\i\in\I_{t-1}} {W_\i} f_{t-1}(X_{\an(\i)})}
   {(n-2)\: \sum_{\i\in\I_{t-1}} {W_\i} \tilde h_{t-1}(X_{\i}) +2\sum_{\i\in\I_{t-1}} {W_\i} \Vert\lik_t h_t\Vert_\infty} \\
 &\geq  \frac{\lam\sum_{\i\in\I_{t-1}} {W_\i} f_{t-1}(X_{\an(\i)})}
   {\lam\: \sum_{\i\in\I_{t-1}} {W_\i} \tilde h_{t-1}(X_{\i}) +2\sum_{\i\in\I_{t-1}} {W_\i} \Vert\lik_t h_t\Vert_\infty} \\
 &=  \frac{\sum_{\i\in\I_{t-1}} {W_\i} f_{t-1}(X_{\an(\i)})}
   {\sum_{\i\in\I_{t-1}} {W_\i} h_{t-1}(X_{\i})}.\\
\end{split}
\end{equation}
It is now enough to apply $\Ex$ to both sides of the above inequality to obtain \eqref{lem:Cak}.

The rest of the proof is easy. By definition of PTGS, using the form of functions $f_t$
and \eqref{lem:Cak} we obtain
\begin{equation}\nonumber
\begin{split}
 \Pr(X'_{1:m}\in \D|X_{1:m}=x_{1:m})&=\Ex\dfrac{\sum_{\i\in\I_m} W_\i \Ind(X_{\an(\i)}\in\D)}{\sum_{\i\in\I_m} W_\i}\\
 &=\Ex\dfrac{\sum_{\i\in\I_m} W_\i f_m(X_{\an(\i)})}{\sum_{\i\in\I_m} W_\i h_m(X_{\i})}\\
 &\geq \Ex\dfrac {f_0}{h_0}=\dfrac{\z\pi(\D)}{h_0}, 
\end{split}
\end{equation}
which concludes the proof.

\subsection{Proof of Theorem 3}

\begin{proof}
As in the proof of Theorem \ref{th:Main}, 
we note that the configuration obtained by
cPTPF has the extended target distribution $\phi$, provided that at the input $X_{1:m}\sim\pi$. 
We are to show that $(\I,\E,\XX,\S)\sim \phi$ implies $(\I,\E',\XX,\S)\sim \phi$.
PTGAS-dt consists of a series of samplings from full conditional distributions of single arrows,
$\phi(\i'\to\j|\I,\E\setminus\{\i\to\j\},\XX,S)$. Formulae \eqref{eq:GenExTarg}
and \eqref{eq:GenExProp} allow us to compute these conditional distributions. 
Crucial points are the following. The weights do not depend on the arrows.
The intensity parameters do not depend on the arrows either, because they are given by 
\eqref{eq:DiscreteLambdas}. The same is true for the estimate $\hat{z}$, because
$\hat z=\sum_{\s\in\It} w_\s /(\prod_{g=0}^{m-1}\lambda_g)$. Consequently,
if  $\I,\xx,\s$ are  fixed and we denote $\E'=(\E\setminus\{\i\to\j\})\cup\{\i'\to \j\}$ then we have 
\begin{equation}\nonumber
 \frac{\phi(\I,\E',\xx,\s)}{\phi(\I,\E,\xx,\s)}=
                        \frac{w_{\i'}\p_{\t-1}(x_{\i'},x_{\j})}{w_{\i}\p_{\t-1}(x_{\i},x_{\j})}.
\end{equation}
Therefore,  $\phi(\E'|\I,\E\setminus\{\i\to\j\},\xx,\s)\propto w_{\i'}\p_{\t-1}(x_{\i'},x_{\j})$. We have shown that
$\t$th small step in  PTGAS-dt, that is sampling new edge $\i'\to\j$, preserves $\phi$. 
\end{proof}
\goodbreak

\subsection{Proof of Theorem 4}

\def\pro{'}
\def\den{k}

Similarly as in the proof of Theorem \ref{th:ASdiscrete} it is enough to consider a small step of sampling an arrow $\i\pro\to\j$.
We are to  show that the move from $\i\to\j$ to $\i\pro\to\j$ preserves $\phi$. Write $\E\pro:=(\E\setminus\{\i\to\j\})\cup\{\i\pro\to \j\}$. 
Below we use lower case letters to denote values of random variables appearing in \eqref{eq:ContinuousLambdas} and in pseudo-code PTGAS-ct.
Let us recall that $\phi$ is given by \eqref{eq:GenExTarg}:
\begin{equation}\nonumber
\begin{split} 
 \phi(\I,\E,{\d}\xx,{\d}\tt,\s)
              &=\frac{1}{\z}\prod_{\k\in\I\setminus\It} \exp\left[-\lambda_{\k}w_{\k}\right] \left(\lambda_{\k}w_{\k}\right)^{|\ch(\k)|}
  \prod_{\l\in\ch(\k)}\den\left(x_{\k},t_{\k}; x_{\l},t_{\l}\right)\d x_{\l}\d t_{\l}\frac{w_{\s}}{\c_{\pa(\s)}}.\\
\end{split}
\end{equation}
We claim that the following equality holds
\begin{equation}\label{eq:AScont}
\frac{\phi(\I,\E\pro,\d\xx,\d\tt,\s)}{\phi(\I,\E,\d\xx,\d\tt,\s)} =
                         \frac{w_{\i\pro}  w_{\i\pro\to\j} \cdot \den\left(x_{\i\pro},t_{\i\pro}; x_{\j},t_{\j}\right)}{\c_{\pa(\i\pro)}} \cdot   
                         \frac{\c_{\pa(\i)}}{w_{\i} w_{\i\to\j} \cdot \den\left(x_{\i},t_{\i}; x_{\j},t_{\j}\right)},
\end{equation}
provided that $\j\in\F^r\cap\an(\S)$ and $\i,\i\pro\in\F^p$ with $p<r-1$. The correctness of PTGAS-ct will follow, because \eqref{eq:AScont} shows that
the probability of sampling an arrow $\i\pro\to\j$ is proportional to  $\phi(\E'|\I,\E\setminus\{\i\to\j\},\xx,\tt,\s)$ in ``strata'' corresponding
to strips $\F^p$. 

We are left with the task of proving \eqref{eq:AScont}. 
Begin with the following simple observations:
\begin{itemize}
\item For any $q$ sets of nodes $\F^{q}$, $\G^{q}$ and  $\F^{q}_\circ$ remain unchanged if $\i\to\j$ is replaced by $\i\pro\to\j$
      (because $\i$ and $\i\pro$ belong to the same strip).
\item Since $\j \in \G^{r-1}$ we ensure that $w_{\j}^{r}$ remains unchanged. The same is true for all descendants of $\j$ and $\i\pro$.
        Consequently for any node $\l$, its Poisson parameter  $w_\l \lambda_\l$ remains unchanged.
 \item Quotient $\den\left(x_{\i\pro},t_{\i\pro};\d x_{\j},\d t_{\j}\right)/\den\left(x_{\i},t_{\i};\d x_{\j},\d t_{\j}\right)$
 appears in \eqref{eq:AScont}, for the obvious reason.
 \item Expression $\lambda_{\i\pro}w_{\i\pro} / \lambda_{\i}w_{\i}$ appears in the ratio, because $|\ch(\i\pro)|$ is increased
 by one and $|\ch(\i)|$ is decreased by one.
\end{itemize}
Apart from the items listed above, we have to consider changes in the expression $\c_{\pa(\s)}=\prod_{\k\in \an(\s)\setminus\{\s\} }\lambda_\k$ 
(we assume that $\j\in\an(\s)$). Let us use the notations $\c_{\pa(\s)}$ and $\c_{\pa(\s)}\pro$ for the quantities before and after replacing
$\i\to\j$ by $\i\pro\to\j$. 

Note that $\c_{\pa(\s)}$ can be factorised into $\c_{\pa(i)} \lambda_\i  \lambda_{\j} \mathit{L}$ where 
$\mathit{L}$ depends only on descendants of $\j$ and consequently cancels out in the ratio $\c_{\pa(\s)}/\c_{\pa(\s)}\pro$.
According to \eqref{eq:ContinuousLambdas},
\begin{equation}\nonumber
 \lambda_\j=\dfrac{1}{w_\j}\dfrac{w_\j^r}{w_\circ^r}b\left(\lambda_0-|\G^r|\right),
\end{equation}
In this expression only $w_\j$ may change if we replace $\i \rightarrow \j$ by $\i\pro \rightarrow \j$ ($w_\j=w_{\i\to\j}$ is replaced by $w_{\i\pro\to\j}$). 
This fact is easy to see from the preceding discussion.  As a result, 
\begin{equation}\nonumber
 \frac{\c_{\pa(\s)}}{\c_{\pa(\s)}\pro} = \frac{\c_{\pa(i)}   \lambda_\i  w_{\i\pro\to\j}}{\c_{\pa(i\pro)}\lambda_{\i\pro} w_{\i\to\j}}.
\end{equation}
Combining everything together we arrive at 
\begin{equation}\nonumber
\frac{\phi(\I,\E\pro,\d\xx,\d\tt,\s)}{\phi(\I,\E,\d\xx,\d\tt,\s)}  = 
\frac{\den\left(x_{\i\pro},t_{\i\pro};\d x_{\j},\d t_{\j}\right)}{\den\left(x_{\i},t_{\i};\d x_{\j},\d t_{\j}\right)} \cdot
\frac{\lambda_{\i\pro}w_{\i\pro}}{\lambda_{\i}w_{\i}} \cdot 
 \frac{\c_{\pa(i)}   \lambda_\i w_{\i\pro\to\j}}{\c_{\pa(i\pro)}  \lambda_{\i\pro} w_{\i\to\j}}.
\end{equation}
We have verified \eqref{eq:AScont} and thus finished the proof.

\section{Comparison of parallel implementations}

In the literature there exist several approaches to construction of enhanced  parallel
SMC algorithms. One of those includes reducing overhead introduced by cumulative weight normalisation. 
This has been a focus of \citep{parallelResampling} -- to this end
they employ methods based on Metropolis and rejection resampling to bypass difficulties caused by vector reduction step.
Authors' scheme enables more efficient GPU implementations, by far exceeding PTPF's capabilities in this area, 
though the amount of data shared between threads is still equal to $\mathcal{O}(N)$ in a worst case scenario.

Both 
algorithms do not avoid the necessity of machine-wide synchronisation after each propagation step -- 
in case of PTPF all threads must receive each other's normalizing constants, in case of \citep{parallelResampling} 
approach all threads must receive unnormalized particles' weights .

PTPF bears a strong resemblance to a scheme introduced in \citep{async} -- "the Particle Cascade".
Authors' approach enables asynchronous particles propagation, in effect getting rid of crude global 
synchronisation after each step
(but synchronisation is still present implicitly because strong ordering on the times of particles reaching next 
propagation step must be imposed)
 and analogously to PTPF considerable decrease of amount of data which must be shared between threads.

PC's unique scheme encourages feeding new particles into system continuously  -- to progressively improve results.
However, fluctuations in number of alive particles are harder to control. 
In comparison in the case of PTPF we have found those alterations to be negligible (see Figure \ref{fig:PopStats}) -- 
a result of more restrictive synchronisation.

\bibliographystyle{abbrvnat}
\bibliography{refs}

\end{document}